\documentstyle[11pt,aaspp4]{article} 

\newcommand{\etal}{et al. }
\newcommand{\kms}{km~s$^{-1}$}
\newcommand{\Lya}{Ly$\alpha$}
\newcommand{\PKS}{PKS~2155-304}    

\begin{document}

\title{ A CLUSTER OF LOW-REDSHIFT LYMAN-$\alpha$ CLOUDS TOWARD \\
PKS~2155-304. I. LIMITS ON METALS AND D/H  } 

\author{J. MICHAEL SHULL, STEVEN V. PENTON, JOHN T. STOCKE, \\
AND MARK L. GIROUX}

\affil{Center for Astrophysics and Space Astronomy, \\ 
Department of Astrophysical and Planetary Sciences,\\ 
University of Colorado, Campus Box 389, Boulder, CO 80309 \\
Electronic mail: mshull@casa.colorado.edu, spenton@casa.colorado.edu, \\
stocke@casa.colorado.edu, giroux@casa.colorado.edu   }

\author{J. H. VAN GORKOM AND YONG-HAN LEE}
\affil{Astronomy Department, Columbia University, \\
538 W. 120$^{\rm th}$ St., New York, NY 10027 \\
Electronic mail: jvangork@astro.columbia.edu }

\author{CHRIS CARILLI}
\affil{National Radio Astronomy Observatory,\\
P.O. Box O, Socorro, NM 87801 \\
Electronic mail: ccarilli@aoc.nrao.edu }

\begin{abstract}

We report observations from the {\it Hubble Space Telescope} (HST)
and the VLA on the galactic environment, metallicity, and D/H in
strong low-redshift \Lya\ absorption systems toward the bright BL~Lac object
\PKS.  These studies are intended to clarify the origin and chemical
evolution of gas at large distances from galaxies.  With GHRS/G160M
data at $\sim20$ \kms\ resolution, we detect a total of 14 \Lya\ absorbers,
six of them clustered between $cz = 16,100$ and 18,500 km~s$^{-1}$.
Although {\it ORFEUS} studies claimed Lyman continuum (Lyc) absorption at 
$z \approx 0.056$ with N(H~I) = $(2-5) \times 10^{16}$ cm$^{-2}$,
the \Lya\ data suggest a range, N(H~I) $= (3-10) \times 10^{14}$ cm$^{-2}$. 
Even higher columns, needed for consistency with the {\it ORFEUS} Lyc
results, are possible if the \Lya\ line core at $17,000 \pm 50$ \kms\ 
contains narrow H~I components.
We identify the \Lya\ cluster with a group of five H~I galaxies offset by
$(400-800)h_{75}^{-1}$ kpc from the sightline.  The two strongest absorption 
features cover the same velocity range as the H~I emission in the two 
galaxies closest to the line of sight.  If the \Lya\ is associated with 
these galaxies, they must have huge halos ($400-500 h_{75}^{-1}$ kpc) of 
highly turbulent, mostly ionized gas. The \Lya\ absorption could also 
arise from an extended sheet of intragroup gas, or from smaller primordial 
clouds and halos of dwarf galaxies.  

We see no absorption from Si~III $\lambda1206$, C~IV $\lambda1548$, or
deuterium \Lya\ at the expected positions of the strongest \Lya\ absorbers.  
Photoionization models yield ($4\sigma$) limits of 
(Si/H) $\leq 0.003 {\rm(Si/H)}_{\odot}$, 
(C/H) $\leq 0.005 {\rm (C/H)}_{\odot}$, 
and (D/H) $\leq 2.8 \times 10^{-4}$ if N(H~I) has the {\it ORFEUS} value
of $2 \times 10^{16}$ cm$^{-2}$.  The limits increase to 0.023 solar 
metallicity and D/H $\leq 2.8 \times 10^{-3}$ if N(H~I) is only 
$2 \times 10^{15}$ cm$^{-2}$.  These limits can be improved
with further studies by HST/STIS and measurements of the
Lyc and higher Lyman series absorption by FUSE.  However, the current data
suggest that the intergalactic gas
in this group has not been enriched to the levels suggested by X-ray
studies of intracluster gas.
Because of their low metallicity and large distance from galaxies,
these absorbers could be primordial gas clouds.   

\end{abstract}

\keywords{cosmology: observations --- galaxies: abundances --- 
    ISM: H I --- quasars: absorption lines }

\normalsize

\newpage

\section*{1. INTRODUCTION }

The origin of the low-redshift \Lya\ absorption clouds discovered by the
{\it Hubble Space Telescope}  (Bahcall \etal 1991; Morris \etal 1991;
Stocke \etal 1995) remains a mystery, made more tantalizing by the 
possibility that they may contain significant mass.   
Are they remnants of the high-redshift Ly$\alpha$ forest, or are they 
associated with past episodes of star formation, galactic outflows, 
and galaxy interactions?  Based on their frequency, estimated size, 
and an ionization correction, they appear to contain a substantial amount
($\Omega_b \geq 0.003 h_{75}^{-1}$) of dark baryons (Shull \etal 1996;
Shull 1997).  There are also strong indications that some of these 
absorbers are associated with galaxies (Lanzetta \etal 1995; Stocke 
\etal 1995; van Gorkom et al. 1996), although the distances to the nearest
bright galaxies are often quite large (Morris \etal 1993;
Stocke \etal 1995; Shull \etal 1996; Grogin \& Geller 1998).  

From numerical models, it now appears that the high-$z$ Ly$\alpha$ forest
is part of a complicated gaseous structure, formed by the gravitational
fluctuations of dark-matter potentials (Cen \etal 1994; Hernquist \etal 
1996; Zhang \etal 1997) and at least partially ``polluted'' by heavy element 
nucleosynthesis during the epoch of galaxy formation.  At high redshift, 
heavy elements (C~IV, Si~IV, C~II) have been detected at $\sim 10^{-2.5}$ 
times solar metallicity in 50--75\% of the Ly$\alpha$ forest clouds above
$10^{14.5}$ cm$^{-2}$ column density (Cowie \etal  1995; Tytler 1995).
From X-ray measurements of Fe-line strengths in galaxy clusters, 
Mushotzky \& Loewenstein (1997) suggest that intragroup gas might be 
enriched to levels of 10\% solar metallicity at $z > 0.4$.  However, 
extrapolating these ideas to low redshift or to all \Lya\ clouds is 
difficult.  Although a few metal-line absorption systems have been detected  
at moderate redshift, no low-redshift \Lya-only absorber has yet been examined 
for the presence of heavy elements, which would indicate contamination by star 
formation.  To investigate this possibility in the \Lya\ forest   
requires finding absorbers of sufficient column density to detect metals 
[N(H~I) $>10^{16}$ cm$^{-2}$] and located appropriately distant 
$(D > 200$ kpc) from neighboring galaxies to suggest primordial gas. 

We now believe to have found an ideal target and absorbers, based on 
our newly analyzed ultraviolet spectra of \PKS\
from the {\it Hubble Space Telescope} (HST) and 21-cm emission images 
from the {\it Very Large Array} (VLA).
Toward this bright, variable ($V = 13.1-13.7$) BL~Lac object, a strong 
\Lya\ line at $cz \approx 17,000$ \kms\ was initially identified by the 
IUE satellite (Maraschi et al. 1988).  This and other \Lya\ lines were 
confirmed by low-resolution HST spectra with the Faint Object
Spectrograph (FOS/G130H) (Allen \etal 1993) and with the Goddard High
Resolution Spectrograph (GHRS/G140L) (Bruhweiler \etal 1993).  
The absorption was resolved by HST into at
least two and possibly three systems.  Using fits to
Lyman-continuum (Lyc) absorption near 960--970 \AA\ in low-resolution 
{\it ORFEUS} spectra, Appenzeller \etal (1995) 
claimed that the 17,000 \kms\ absorbers had a combined column 
N(H~I) $\approx (2-5) \times 10^{16}$ cm$^{-2}$.  From our HST/GHRS 
measurements of the corresponding \Lya\ absorption, we estimate 
a range, N(H~I) $= (3-10) \times 10^{14}$ cm$^{-2}$.   Later in this paper, 
we will assess the accuracy of the H~I column densities associated 
with the Lyc absorption and discuss the conditions
under which the {\it ORFEUS} and HST measurements could be reconciled.  

Previously, our group observed the \PKS\ field with the
VLA (van Gorkom \etal 1996) and found evidence for 
three H~I galaxies corresponding to \Lya\ absorbers at 
$cz = 5100$, 16,500, and 17,000 \kms.  
We associated the 17,000 \kms\ absorbers with a small group of 
galaxies offset from the \PKS\ sightline by $(400-800) h_{75}^{-1}$ kpc
for a Hubble constant $H_0 = (75$~km~s$^{-1}$~Mpc$^{-1}) h_{75}$.  
In this paper, we combine new HST and VLA observations 
with theoretical interpretation of the 
\PKS\ sightline, which is unusual in the large number of strong \Lya\
absorbers and their association with four large galaxies
located within $\sim1$ Mpc. This number of galaxies would be
high by chance, based upon the observed two-point correlation function. 
The average space density of galaxies in this local region, $n_{\rm gal} 
\sim 1$ Mpc$^{-3}$, is $\sim100$ times the large-scale
average density of $L_*$ galaxies (Marzke, Huchra \& Geller 1994). 
This suggests the presence of a small group that has recently turned 
around from the Hubble flow, much  like our Local Group.  These galaxies 
probably have not undergone significant mergers, but the presence 
of strong, broad \Lya\ absorption from extended gas may indicate some 
dynamical interaction. 

In \S~2.1, we describe new HST/GHRS observations at 20 \kms\ resolution
between 1258 and 1293 \AA\ ($cz = 10,440 - 19,060$ \kms). 
We detect 7 \Lya\ absorbers with equivalent widths ranging
from 68 to 467 m\AA.  We also reanalyze an archival GHRS/G160M spectrum at
1223 -- 1258 \AA\ ($cz = 1809 - 10,446$ \kms) 
and identify 7 definite \Lya\ absorbers (21--201 m\AA).   
In \S~2.2, we present new VLA studies of H~I emission in the 2155-304 field 
from $cz = 16,283 - 17,571$ \kms, a velocity range that includes four 
\Lya\ absorption systems.  In \S~3, we discuss the results of 
these studies.  The new VLA data provide accurate correspondence
with the \Lya\ absorbers, suggesting that the absorption
might arise from extended halos or intragroup gas over a region 1 Mpc 
in diameter that could total $10^{11} - 10^{12}~M_{\odot}$ in mass,
if bound.  
The HST data also allow us to set limits on the metallicity 
of the strongest absorbers from the absence of Si~III $\lambda1206$ and 
C~IV $\lambda1548$, and they provide a limit on D/H from residual (D~I) 
\Lya\ absorption in the shortward wing of the 
strongest absorber. In \S~4 we summarize our conclusions and give
suggestions for further study.  Somewhat contrary to the kinematic
evidence, the low metallicities, [Si/H] $\leq 0.003$ solar and 
[C/H] $\leq 0.005$ solar, would suggest that some primordial gas 
may still reside amidst the large-scale filaments of galaxies.

\section*{2. OBSERVATIONS }

\subsection*{2.1. New HST/GHRS Spectra }

The target, \PKS, lies at redshift $z = 0.116$ or $cz = 34,775$ \kms\
(Falomo, Pesce \& Treves 1993).  It was observed by HST during Cycle 6 
on October 5, 1996, using the GHRS with the G160M grating
and post-COSTAR optics.  The continuum flux near 1280 \AA\  
was $(8.7 \pm 0.3) \times 10^{-14}$ ergs cm$^{-2}$ s$^{-1}$ \AA$^{-1}$, about
70\% of the median flux for this object over the past 15 years
of observations in the IUEAGN database (Penton, Shull \& Edelson 1998).  
Over the interval 1258--1293 \AA, we obtained a resolution of 4.5 \kms\ 
per 0.018 \AA\ quarter-stepped pixel or $\sim20$ \kms\ per resolution
element.  

The wavelength scale was determined by assuming that the Galactic
interstellar S~II absorption features at 1250.584, 1253.811, and
1259.519 \AA\ lie at zero velocity in the local standard of rest (LSR).
Based on the Bell Labs 21-cm survey
(Stark \etal 1992), the dominant H~I absorption in this direction
lies at $V_{\rm LSR} \approx V_{\rm hel}$, within the accuracy
($\pm 2$ \kms) of the observations.  This correction in the
wavelength scale in the new data (Fig. 1) was $+0.029$ \AA, and that
in the archival data (Fig. 2) was $-0.001$ \AA.  Our observations  
correspond to redshifted velocities $cz = 10,440 - 19,060$ 
\kms\ ($z = 0.035 - 0.064$) in the \Lya\ line. According to previous 
convention, we quote heliocentric velocities, $cz$, and compute expected 
line positions from $\lambda = \lambda_0 (1+z)$.  We do not make relativistic
corrections to the velocities.   

Figure 1 shows our new GHRS/G160M data (1258--1293 \AA) with
S/N $\approx 20$. Figure 2 shows our reanalysis of an unpublished 
archival GHRS/G160M spectrum with S/N $\approx 34$ 
(1223--1258 \AA) taken with pre-COSTAR optics.  
The 17,000 \kms\ \Lya\ absorbers are the strongest 
of the low-redshift ($z < 0.1$) absorbers found in HST searches 
other than associated absorbers with $z_{\rm abs} \approx z_{\rm em}$. 
Because of the exceptional brightness 
of the background source ($\sim13$ mag) and the high H~I column densities 
($\geq10^{16}$ cm$^{-2}$), these absorbers are ideally suited for measuring 
heavy-element abundances in the low-$z$ forest.  As of this writing,
our group has studied $\sim 100$ low-redshift \Lya\ absorbers toward 11 bright 
quasars and Seyfert galaxies (Stocke \etal 1995; Shull \etal 1996; Shull
1997) with more absorbers under analysis (Penton \etal 1998).  Our 
goal is to understand the physical structure of these clouds 
and their possible connections with galaxy halos, large-scale structure, 
and voids.  Our data reduction method was described in these
earlier papers.  Complete results on the seven targets in our HST Cycle 6 
observations, together with calibration, are described in 
Penton \etal (1998).  

The GHRS spectra were taken through the $2''$ large science aperture, 
using the standard quarter-diode sub-stepping pattern to yield pixels of
0.018 \AA\ in FP-split mode.  
We recalibrated our spectra using IRAF/STSDAS/CALHRS and the final GHRS
reference files (Sherbert \& Hulbert 1997) with polynomial background 
subraction.  One noteworthy point concerns our treatment of the HST/GHRS
error vectors, which affect how we gauge the
statistical significance of weak features.  We found that the
error vectors produced by the IRAF/STSDAS program {\bf specalign} in the
HST Data Handbook do not agree with those obtained by standard
propagation of errors from individual subexposures.   
We have therefore chosen to perform our own error propagation and
spectral coaddition using the IDL software package.  Spectral coaddition 
was weighted by exposure time on a quarter-stepped, pixel-by-pixel 
basis, removing blemished pixels.  This procedure gives some pixels 
less exposure time than others, but it prevents photocathode blemishes 
from being erronously flagged as absorption features. In some
cases, this procedure results in no exposure time being available for
some pixels; straight lines between the last known good flux values are 
then used in Figures 1 and 2.

\begin{figure}
   \epsscale{0.70}
   \plotone{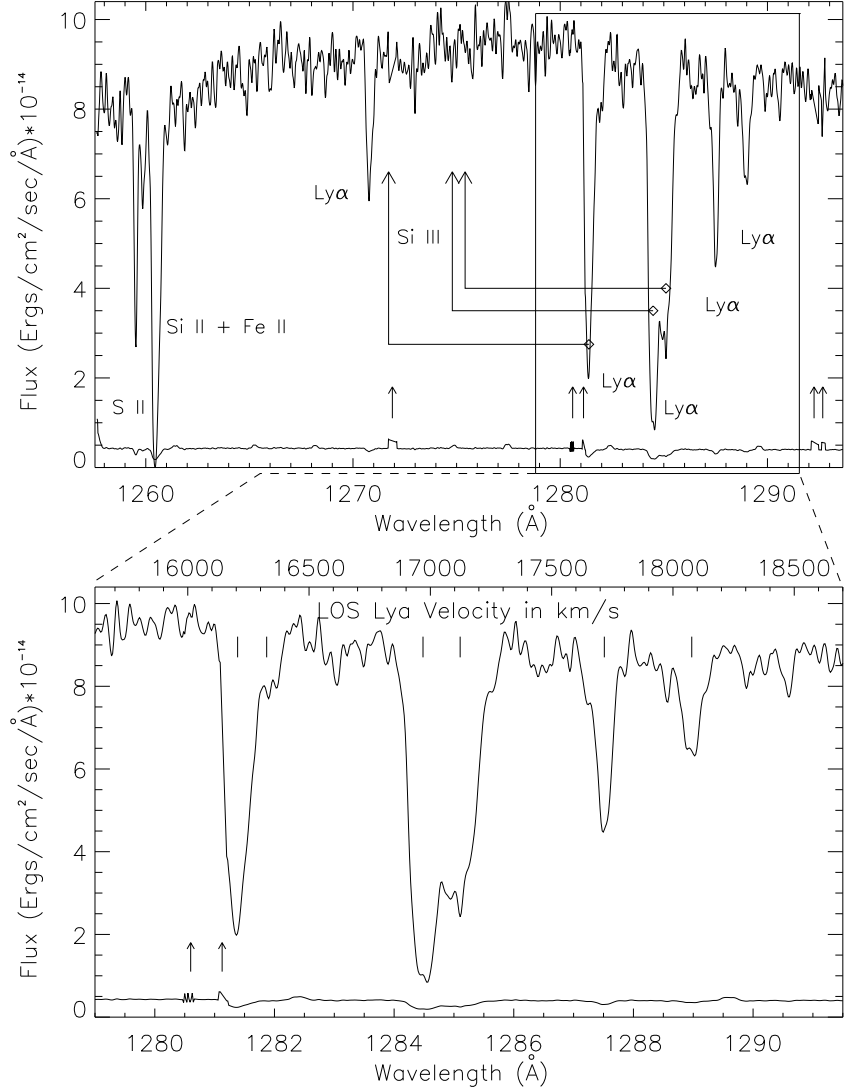}
   \caption[]{{\bf Top:} Smoothed HST/GHRS (G160M) spectrum of PKS~2155-304 
    shows multiple \Lya\ absorption systems, with several strong absorbers
    at $cz \approx 17,000$ \kms. Because of smoothing, the absorbers do not
    reach zero flux; error vector ($1 \sigma$) is shown at base. Small arrows 
    near error vector show deleted photocathode blemishes.  Lines at 
    1259--1261~\AA\ are Galactic interstellar absorption; the weak line 
    at 1259.869 \AA\ is Si~II in a high-velocity cloud at
    $V \approx -132$ \kms.  Upper limits on Si~III $\lambda1206.50$ 
    absorption at 1274.7~\AA\ and 1275.2~\AA\ (see arrows) correspond to 
    [Si/H] $< 0.003$ solar abundance ($4 \sigma$).  {\bf Bottom:} Six \Lya\ 
    absorbers ($cz = 16,120 - 18,590$ \kms) including strong features near 
    1281 and 1285 \AA, estimated to have total N(H~I) $= (2-5) \times 10^{16}$ 
    cm$^{-2}$ from Lyc absorption (Appenzeller et al. 1995). } 
\end{figure}

\begin{figure}
 \epsscale{0.90}
  \plotone{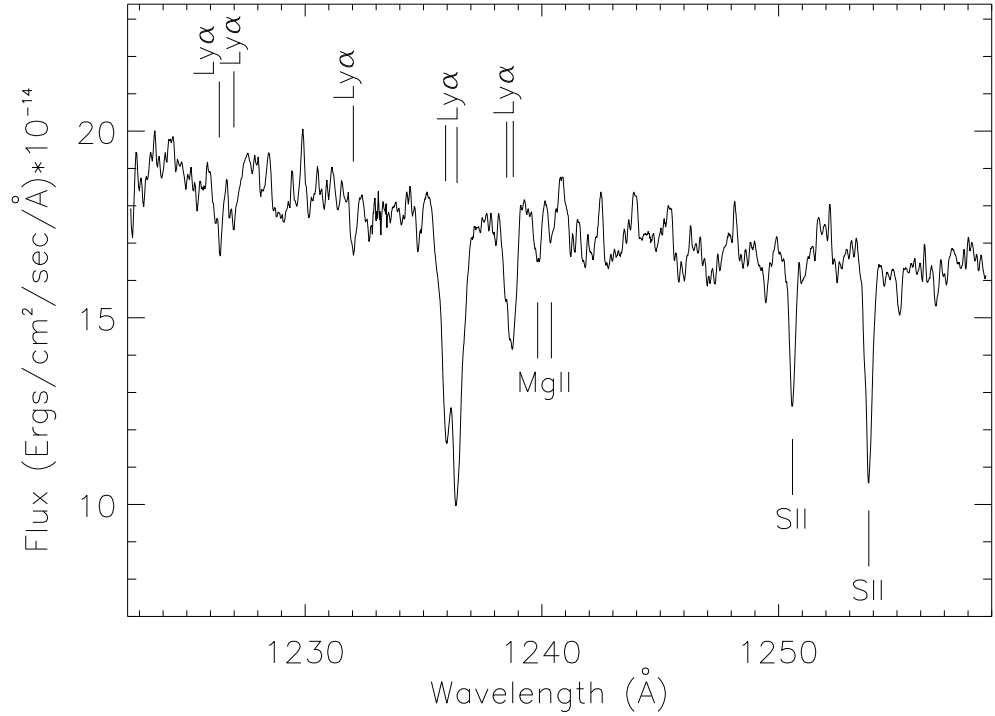}
  \caption[]{Reanalyzed archival GHRS/G160M spectrum from 1223 -- 1258 \AA\
   shows 7 \Lya\ lines and Galactic interstellar lines of S~II (1250.584, 
   1253.811~\AA) and Mg~II (1239.925, 1240.395~\AA). The \Lya\ feature
   near 1236~\AA\ consists of two components separated by
   122 \kms\ at comparable velocity to an H~I galaxy at $cz \approx 5100$ 
   \kms\ (van Gorkom \etal 1996).  We identify the blended absorbers at 
   1238.426 and 1238.744 \AA\ as \Lya, despite their proximity to 
   Galactic N~V (see text).  }
\end{figure}

The significance of the detected lines (in $\sigma$) is defined as 
the integrated S/N per resolution element of the fitted absorption 
feature.  The status of ``definite absorber'' is given to features
greater than $4\sigma$ significance (Penton \etal 1998).  
Table 1 lists the identified absorption features and their equivalent 
widths.  We see the expected Galactic interstellar lines
of S~II $\lambda1259.519$ and Si~II $\lambda1260.422$, as well as
a high-velocity cloud in Si~II at $V_{\rm LSR} = -132$ \kms,
which was also seen in C~IV at $V_{\rm LSR} = -141 \pm 9$ \kms\ 
by Sembach \etal (1998).  We also confirm the 
strong intergalactic \Lya\ absorbers near 17,000 \kms.  A blowup 
of the 1278--1292 \AA\ region shows that many of the \Lya\ lines 
have velocity substructure, which we model as separate Gaussian components. 
We included these additional components to reduce the $\chi^2$
per degree of freedom whenever the resulting components were
above $4\sigma$ significance.  In no case did we allow more than
two components per blended absorption feature, although we occasionally
tested for three.

In addition to the definite \Lya\ absorbers, we found a number
of features at the $3-4 \sigma$ level that might be verified
with future STIS data. Examples of these ``possible features''
occur at 1229.0, 1243.2, 1249.5, 1264.8, 1286.45, and 1290.58 \AA.  
The local continua are uncertain near these features, and some
of the features are too broad ($\geq 120$ \kms) to be discrete 
\Lya\ absorbers by our criteria.  In these cases, we have decided 
to be conservative and not classify them as definite.

We also analyzed an archival spectrum (Fig. 2) taken with the
GHRS/G160M using pre-COSTAR optics. This spectrum was reduced and analyzed
in a manner identical to that in Figure 1, with the wavelength 
scale zero point set to the LSR using the Galactic interstellar S~II 
lines at 1250.584 and 1253.811 \AA.  Between 1223 and 1258 \AA, we detect
7 definite \Lya\ lines and three Galactic interstellar lines: the
two S~II resonance lines and the weak Mg~II doublet.  

The strong \Lya\ feature at 1236 \AA\ is easily detected
and is resolved into two subcomponents separated by $122$ 
km~s$^{-1}$. This pair of lines, at $cz = 4999$ and 5121 km~s$^{-1}$,
lies at the same velocity as the H~I detected galaxy at
$21^{\rm h}~57^{\rm m}~04^{\rm s}$, $-30^{\circ}~25.5$
just off the eastern edge of Figure 3 and discussed in detail by
van Gorkom \etal (1996).  This sub-$L_*$ galaxy is located
$\sim300h_{75}^{-1}$ kpc from the \PKS\ sightline. 
The absorption line at 1238.744~\AA\ lies
0.077 \AA\ blueward of the expected position of Galactic N~V
$\lambda1238.821$. We identify this line as \Lya\ because of the 
slight wavelength offset and what would be an unusual strength
for N~V.  Based on the absence of the weaker line of the N~V doublet at 
1242.804~\AA, we believe that N~V $\lambda1238.821$ contributes 
at most 30\% (at the $4\sigma$ level) to the equivalent width of the 
\Lya\ feature at 1238.744 \AA.

The \PKS\ sightline appears to be unusual, both in the number of \Lya\ 
absorbers and in their strength.  In total, from 1223 -- 1293 \AA, we 
identify 14 \Lya\ absorbers with significance greater than $4 \sigma$.  
With a correction for regions blocked by Galactic S~II and Si~II, these 14 
\Lya\ absorbers correspond to an uncorrected frequency 
$d {\cal N}/dz = 250$ above 21 m\AA\ (N$_{\rm HI} \geq 10^{12.6}$ cm$^{-2}$),
for N$_{\rm HI} \geq 10^{13}$ cm$^{-2}$. This frequency is somewhat
higher than the mean value, to similar absorption strength,
in sightlines studied with the GHRS during cycles 2 and 4 (Shull et al.
1996).   Probably the most unusual
aspect of this sightline, however, is the large number of
strong \Lya\ absorbers near 17,000 \kms. 
The strong correspondence in recession velocity between these
\Lya\ absorbers and the surrounding galaxies argues that these
clouds are not ejected from the BL~Lac object.  Rather, they
appear to be intervening clouds at distances given by the
Hubble law, $d \approx (227~{\rm Mpc})(V/17,000~{\rm km~s}^{-1})h_{75}^{-1}$,
as we assume throughout this paper.

\subsection*{2.2. Deep VLA H~I Imaging}
 
\begin{figure}
   \plotone{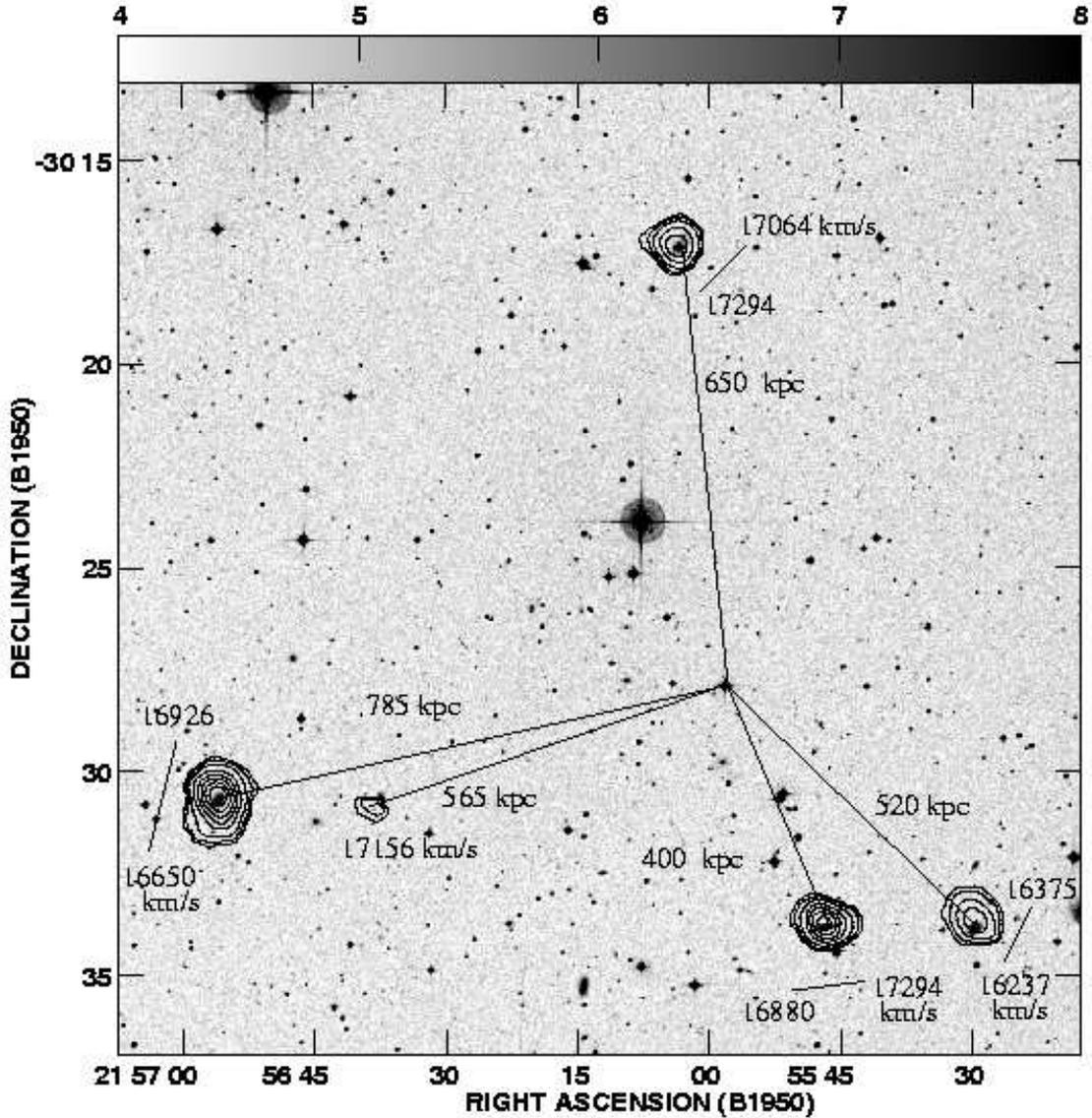}
    \caption[]{An overlay of the total H~I emission  (contours) toward \PKS\
    in the velocity range (16,283 -- 17,571 \kms) on an image of the
    digitized POSS. Five galaxies are detected in H~I at projected distances
    from the sightline toward \PKS\ of (400--785)$h_{75}^{-1}$ kpc.
    Contours levels are 1.65, 3.3, 6.6, 13.2, 19.8, 26.4, 33.0,
    39.6 $\times 10^{19}$ cm$^{-2}$. The image has been corrected for the
    primary beam response. Projected distances from the sight line are
    indicated in $h_{75}^{-1}$ kpc. For each of the galaxies, the velocity
    range and approximate major axis of the H~I emission are indicated. 
    Note that the galaxy closest
    to the sightline covers in H~I emission exactly the same velocity range
    as the broad Ly$\alpha$ absorption seen near 17,000 \kms. The galaxy
    to the southwest covers about the same velocity range as the absorption
    feature near 16,200 \kms.}
 
\end{figure}

Our previous results (van Gorkom \etal 1996)
showed that two of the strongest Ly$\alpha$ absorbers 
found toward PKS~2155-304 at 17,100 \kms\  are located
within a loose group of galaxies. Three galaxies were detected in H~I,
two at the velocities of the absorbers, but no individual galaxy could
be identified as being {\it associated} with the absorber. The high
column densities observed in the absorbers suggested that there may be
an extended component of intergalactic neutral gas. Column densities 
N(H~I) $< 10^{19}$ cm$^{-2}$ are rarely detected in emission
(van Gorkom 1993), and common lore suggests that it may be ionized by the
intergalactic UV background. The Ly$\alpha$ results suggested that this
group may be the ideal place to look for diffuse H~I at very low column
densities. We therefore reobserved the group in hope of detecting
a low surface brightness diffuse H~I component.

The observations were made in 1996 May with the VLA in the 1~km (D) array
with an extended north arm (3~km) to compensate for the low declination
of the source. The total integration time was 40 hrs spread over 8
different runs.  All observations were centered at the radio position of
PKS~2155-304, $21^{\rm h}~55^{\rm m}~58.30^{\rm s}$,
$-30^{\circ}~27'~54.4''$ (B1950). 
We used a total bandwidth of 6.25 MHz centered at 16,880 \kms; the usable 
velocity range is about 1300 \kms.  On-line Hanning smoothing was employed, 
after which every other channel was discarded, leaving a set of 31 independent 
channels and resulting in a velocity resolution of 46 \kms. The BL~Lac 
object is a radio continuum source with a variable flux density. We 
measured a flux density of 0.45 Jy at 1.4 GHz. Extreme care was taken to 
properly calibrate the bandpass. Every 25 minutes a bandpass calibration 
was done, and the bandpass solution for these individual scans was 
interpolated to do the correction. As a result, our observations are 
limited by noise rather than by spectral dynamic range, which
is better than 3000:1. 

The U-V data for each of the 8 days were calibrated
independently and inspected for interference and residual calibration
errors. Subsequently, the U-V data of all runs were combined. The continuum
was subtracted by making a linear fit in frequency to the calibrated
complex visibilities of the line free channels (2-5 and 23-25). The resulting
data were clipped at a level of 0.7 Jy to remove man-made and solar
interference. Images were made using a taper that compromised between
optimal surface brightness sensitivity and maximal sidelobe suppression
(using the task IMAGR in AIPS, with robustness factor 1), resulting
in a synthesized beam of $54.4'' \times 38.6''$.
The rms noise in the channel images is 0.15 mJy beam$^{-1}$. 
The instrumental parameters are summarized in Table 2.  

To search for H~I, we imaged the entire primary beam ($1^{\circ} \times  
1^{\circ}$).  To calculate H~I masses, we use the luminosity distance 
(Sandage 1975) assuming that the group is at $z = 0.057$ and using 
$H_0 = 75h_{75}$ \kms~~Mpc$^{-1}$ and $q_0 = 0.5$.
Our 6 $\sigma$ H~I mass limit in the center of the field is 
$(5 \times 10^8~M_{\odot}) h_{75}^{-2}$. At full resolution, the column 
density sensitivity is $2 \times 10^{19}$ cm$^{-2}$ ($5 \sigma$ over 57 kpc 
$\times$ 34 kpc $\times$ 46 \kms). The data were smoothed spatially and in 
velocity down to a resolution $2' \times 3'$, resulting in a detection limit
of $4 \times 10^{18}$ cm$^{-2}$ (5 $\sigma$ over 126 kpc $\times$
189 kpc $\times$ 92 \kms).  Owing to a lack of short spacings, the 
observations are much less sensitive to H~I emission that is completely 
smooth on scales larger than $15'$ in a single 46 \kms\ velocity channel. 
These values are valid for the center of the field. Farther from the center,
they have to be corrected for the change in primary beam response.   
The primary beam pattern is roughly gaussian with a FWHM of $30'$, and
the limits are a factor two worse $15'$ from the field center. 

The properties of the five galaxies detected in H~I are summarized
in Table 3 and shown in Figure 3.  
Perhaps the most interesting result of these observations is what we do not
find. We do not detect a faint intergalactic component in H~I, nor do we
detect any diffuse extended H~I associated with the galaxies. Compared
to van Gorkom \etal (1996), our new data are of much better quality and
only show more clearly that the H~I emission is confined
to the galaxies at column densities well above $10^{19}$ cm$^{-2}$.
In addition to the 3 galaxies detected in the previous observations, we 
detect to the southwest an S0 galaxy
cataloged in the APM catalog (Loveday 1996) with a velocity range partly
outside our band. We see H~I to the northwest at 16,375 \kms, moving toward
the center of the galaxy at lower velocities. 
The lowest velocity where we have usable, though very noisy data, is at
16237 \kms. Since this emission peaks only slightly to the northwest of
the center, we infer that the systemic velocity of the galaxy must be about
16200 \kms. This galaxy was not detected in our previous observations,
because in those data that velocity range was seriously affected by
interference.  

The other
new detection is a tiny dwarf galaxy to the east at 17,156 \kms. In the current
data, the galaxy only shows up in one channel, but at the 8 $\sigma$ level.
Going back to our previous data, we found the dwarf just above the noise
in our oldest data set (and at the edge of the band there). Those data were
taken with 11 \kms\ velocity resolution, and the emission seems to cover
about 60 \kms. An overlay of just the dwarf on the digitized POSS is shown
in Figure 4.  A hint of some faint light ($m_B = 22 \pm 0.5$) can be seen 
on the blue POSS image, but it is clear from the image that similar dwarfs 
could easily be missed optically. In H~I the dwarf is just above our detection 
limit.  Although somewhat brighter galaxies with larger H~I masses could 
have been detected closer to the sightline, in fact none was seen (Fig. 3).  

\begin{figure}
  \plotone{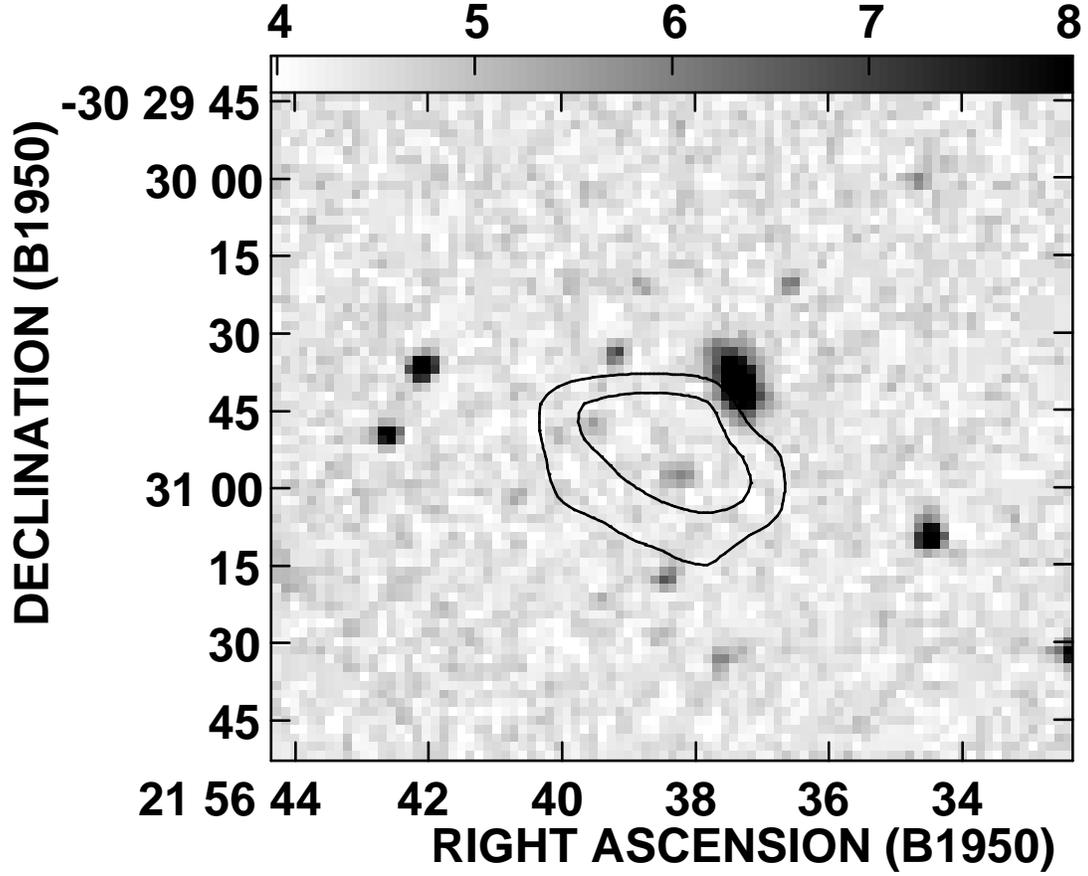}
  \caption[]{Overlay on the POSS of the dwarf galaxy to the east
  of \PKS\ sightline at 17,156 \kms.  Contours are same as in Fig. 3. 
   This galaxy has an estimated
  blue magnitude $m_B = 22 \pm 0.5$.  At the observed redshift distance, 
  $230 h_{75}^{-1}$ Mpc, its absolute magnitude is $M_B \approx -15$,
  or $L \approx 0.014 L_B^*$. }

\end{figure}
 
Are we getting any closer to identifying galaxies associated
with the \Lya\ absorption? There are 3 galaxies with H~I emission
in the range 16,900 -- 17,100 \kms, the velocity of the deepest 
\Lya\ absorption.
Thus, in the first instance, it is not obvious that the absorption would be
{\it associated} with any one of the galaxies.
In our previous work (van Gorkom et al. 1996) we noticed a curious
phenomenon:  in all cases, the \Lya\ absorption was at the systemic
velocity of the closest galaxy. In the current, much improved, \Lya\ data
we notice something even more curious: the two strong 
features at 17,100 \kms\
and at 16,200 \kms\ cover exactly the same velocity range as the H~I emission
in the two nearest galaxies. In Figure 3 we show the total
H~I emission contours overlaid on an optical image (greyscale), and we also
indicate for each galaxy the sense of rotation and velocity extent of
the H~I emission. As in our previous data, there is no evidence for \Lya\
corotating with the H~I disks. The broad widths of the \Lya\ absorbers
in the current data is intriguing. It seems unlikely that the two strong
absorbers are associated with their nearest galaxies only. If the width
of the \Lya\ lines were to reflect the potential of the nearest galaxies, each
would have a halo characterized by a flat rotation curve extending out
to 400 and 500 kpc respectively.
Perhaps more likely, the widths could indicate that the clouds
probe a group potential. In that scenario we may be viewing two
groups along the line of sight, one centered at 17,100 \kms, the other
at 16,200 \kms. The detection of 3 galaxies with systemic velocities
close to 17,100 \kms\ makes it plausible that there is indeed a loose group at
this velocity. Our velocity coverage does not extend to sufficiently
low velocities to make any statements about the presence of a group 
at 16,200 \kms.

\section*{3. RESULTS AND DISCUSSION } 

\subsection*{3.1. Extended Halo and Intragroup Gas }

The combined HST and VLA data suggest that the 17,000 \kms\ 
\Lya\ absorption lines arise in a group environment.  Between
16,100 and 18,100 \kms, we observe 6 \Lya\ absorbers 
and four large (VLA) galaxies at similar velocities.
The velocity centroid and radial velocity dispersion of the 
four galaxies are $\langle V_{\rm gal} \rangle  = 16,853$ \kms\ and
$\sigma_{\rm gal} = 325$ \kms.  The equivalent statistics for
the 6 \Lya\ absorbers are $\langle V_{Ly} \rangle = 17,070$ 
\kms\ and $\sigma_{Ly}  = 740$ \kms. The fact that the strongest \Lya\ 
absorbers (1284.484 \AA\ and 1285.097 \AA) lie close 
to $\langle V_{Ly} \rangle$ 
suggests that the center of the gravitational potential well may be 
centered at $V \approx 17,000$ \kms.  Before we interpret the larger 
velocity dispersion of \Lya\ lines, it is worth noting that this galaxy 
group was observed in H~I emission with the VLA over a more limited 
velocity range (16,283 -- 17,571 \kms) than the \Lya\ lines.  
 
The simplest interpretation of the data is that the absorbing gas arises 
in intragroup material stripped or blown out of the gas-rich galaxies.  
However, there is no direct evidence for this interpretation in 
the H~I emission data.  Nor are galaxy H~I halos expected to be
this dense at such enormous radii (Dove \& Shull 1994).   
Combining the 740 \kms\ velocity dispersion of the 6 definite \Lya\ absorbers 
with the mean projected separation, $\langle R \rangle \approx 1.0$ Mpc 
between the four galaxies, we obtain a virial mass estimate,
\begin{equation}
    M_{\rm vir} \approx  \frac {3 \sigma_r^2 \langle R \rangle } {2G}   
         \approx (2 \times 10^{14}~M_{\odot} )  h_{75}^{-1} \; . 
\end{equation}
The estimated velocity dispersion and size/mass scale are those  
of a modest group, less than the size of the Virgo cluster.  
The mean distance of these galaxies at $\langle V_{\rm gal} \rangle =
16,853$ \kms\ is $225 h_{75}^{-1}$ Mpc or $(m-M) = 36.8$. 
The break in the standard luminosity function is $L_* = (9 \times 10^9 
L_{\odot}) h_{75}^{-2}$ or $M_* = -19.4 + 5 \log h_{75}$ (Marzke \etal
1994) in the Zwicky (blue) magnitude system.  Thus, an $L_*$ galaxy 
in this group would lie at $m_B \approx 17.4$.  Using two slightly 
brighter galaxies in the field for calibration of the SRC-J plate 
magnitudes (Lauberts \& Valentijn 1989), we have estimated
B$_0$ magnitudes for the five galaxies around \PKS.  In terms of
$L_*$, they are:  north ($0.3 L_*$), east ($0.9L_*$), south ($1.4 L_*$),
southwest ($3.0L_*$), and dwarf ($0.014 L_*$).  Thus, three of the 
H~I galaxies have luminosities essentially at or above $L_*$.  

Although drawing statistical inferences from only four galaxies is
risky, we have attempted to estimate the possible number of 
fainter galaxies, both observationally and theoretically. 
In a CFHT image (Wurtz \etal 1997) centered on \PKS, of radius 
$160 h_{75}^{-1}$ kpc at $cz = 17,000$ km~s$^{-1}$, we found five 
galaxies in excess of the background to a limit of $m_B \approx 21$, 
about 3.5 magnitudes below $L_*$.  Over the larger field of 
$600 h_{75}^{-1}$ kpc radius, defined by 
the four H~I galaxies, this corresponds to $\sim60$ excess galaxies. 
However, this photometry does not constrain the redshift of these
excess galaxies; most, if not all, could lie at the redshift,  
$z_{\rm em} = 0.116$, of \PKS\ (Falomo \etal 1993).  Since BL Lac
objects are routinely found in poor clusters (Wurtz \etal 1997),
these five excess galaxies are likely to be associated with \PKS. 

If we integrate a Schechter luminosity function, 
$\phi(L) = \phi_* (L/L_*)^{-1} \exp(-L/L_*)$, down to $0.04L_*$
($m_B \approx 21$), we would expect between 3 and 10 times more 
galaxies, compared to those above $0.3L_*$ and $0.9L_*$, respectively.  
Scaling from the observed H~I galaxies above those limits, we  
would expect to find between 10 and 30 galaxies associated
with this group, down to $0.04L_*$.
Therefore our preliminary optical survey is consistent with 
these expectations. However, since our radio H~I survey 
detected only one dwarf galaxy in the field, gas-rich dwarfs 
may be scarce in this group. We are conducting a spectroscopic 
survey of the \PKS\ field down to $m_B \approx 19$ to test our
hypothesis that the large VLA galaxies are accompanied by
smaller galaxies, some closer to the \PKS\ sightline.

We also investigated the possibility that hot gas associated with the
strong absorbers at 17,100 \kms\ might be responsible
for the 600 eV absorption feature seem by the objective grating
spectrometer on the {\it Einstein Observatory} (Canizares \& Kruper 1984)
and by the BBXRT spectrograph (Madejski et al. 1998).  This possibility
would only work if the X-ray absorption arose from an O~V K-edge
(626 eV rest frame, 592 eV observed) or an O~VIII K$\alpha$ line
(653 eV rest frame, 618 eV observed).  We can rule out the O~V hypothesis
($T \approx 2 \times 10^5$~K) because it would require an unreasonably
large hydrogen column, N$_{\rm H} = (1.7 \times 10^{22}$~cm$^{-2})
\zeta_{\rm O}^{-1}$, for oxygen metallicity $\zeta_{\rm O}$,
in order to produce the observed optical depth, $\tau \approx 2.5$.
The resulting gas densities would be quite high, across a 300 kpc slab, and
the cooling time would be less than $10^5$~yrs.
Line absorption by O~VIII would be more feasible, requiring a column density
N$_H \approx (1.3 \times 10^{19}~{\rm cm}^{-2})\zeta_{\rm O}^{-1}$.
However, neither of these cluster hot-gas models would explain the observed
broad ($\sim30,000$~\kms) absorption widths. We conclude that the
600 eV absorption must be produced elsewhere, probably within \PKS.

Because the four bright galaxies are H~I-rich spirals, it is also 
possible that these galaxies are not bound,  based upon recent 
spectroscopic work on similar galaxy groups (Mulchaey \& Zabludoff 
1998; Zabludoff \& Mulchaey 1998). Even if they are bound, 
these galaxies may not have virialized or 
closely interacted. If this is the case, the \Lya\ clouds may also be
non-virialized, and our estimate of $M_{\rm vir}$ would be too large. 
This interpretation suggests that these
clouds are portions of the gaseous filament out of which these spiral
galaxies formed. If these clouds and galaxies are portions of a bound
group, then we would expect the clouds to have metallicities of 
0.1 -- 0.3 solar, similar to the stripped and virialized gas in rich 
clusters and elliptical-rich groups (Mushotzky \& Loewenstein 1997).  
In the unbound case, not only would X-ray emission be unlikely, but the 
metallicity of the gas would be substantially lower than 10\% solar
(see \S~3.2).

In our earlier papers on the environments  
of the low-$z$ absorbers (Shull \etal 1996; van Gorkom \etal 1996),
we estimated a space density, $\phi_0 \approx (0.7~{\rm Mpc}^{-3})
R_{100}^{-2} h_{75}$, of low-$z$ \Lya\ clouds with
N(H~I) $\geq 10^{13}$ cm$^{-2}$ and characteristic size 
scale $(100~{\rm kpc})R_{100}$,     
in order to explain their frequency per unit redshift. 
This space density is $\sim40$ times larger than that of
$L_*$ galaxies, but comparable to that of dwarf galaxies
with $L \approx 0.01L_*$, suggesting a possible connection.
However, the H~I absorption cross sections of these dwarfs are 
uncertain, and it is unclear whether the extended gas
was stripped out, blown out, or existed primordially.

The main results of our comparison between the \Lya\ absorption and the 
21-cm emission can be summarized as follows: 
(1) 21-cm emission has been detected in this region, 
spread over 1 Mpc of sky and 1300 \kms\ of velocity;
(2) the large distances to the nearest bright galaxies
(400-800$h_{75}^{-1}$ kpc);  and (3) the agreement in velocity
(within $\pm100$ \kms) between several of the \Lya\ absorbers and 
the H~I galaxies to the south, southwest, and north.
The proper interpretation of the \Lya\ absorbers toward \PKS\ 
hinges on several geometrical issues. Does the \Lya\ absorption 
occur in a smoothly distributed layer, 200-600 kpc in depth,
or in smaller, denser clumps? Some evidence for the latter interpretation 
comes from the fact that the \Lya\ absorption occurs in discrete 
systems in velocity space.  Less certain is whether the absorbers 
are kinematically associated with the large galaxies. Although a 
single sight line through the region is not necessarily typical,
the VLA observations suggest that the absorbing gas is distributed 
across a region $\sim1$ Mpc in diameter.  As a first approximation, 
we will model the absorption as a homogeneous slab.  We then explore 
more complex distributions, either in intergalactic clouds
or in dwarf-galaxy halos.   

In the homogeneous approximation, the total mass of gas in the vicinity 
of the cluster of galaxies around \PKS\ can be estimated by assuming a 
slab of radius 500 kpc, depth $D = 400$ kpc, mean column density N(H~I) 
$= 2 \times 10^{16}$ cm$^{-2}$, exposed to an ionizing radiation field with 
specific intensity $I_0 = 10^{-23}$ ergs~cm$^{-2}$ 
s$^{-1}$ Hz$^{-1}$ sr$^{-1}$ at 1 Ryd and spectral slope $\alpha_b = 1.8$, 
as described in \S~3.2. (The BL Lac, \PKS\ cannot dominate the ionizing
flux incident on these clouds, unless its actual redshift is much less than 
the value, $z = 0.116$, from Falomo \etal 1993.)   
If $I_0$ and $D$ are
constant, then N(H~I) $\propto n_{\rm HI} \propto n_H^2$, so that
total density, $n_H$, and gas mass, $M_{\rm gas}$, scale as
[N(H~I)]$^{1/2}$. For the parameters above, the mean densities are
$\langle n_{\rm HI} \rangle = 2.2 \times 10^{-8}$ cm$^{-3}$ and
$\langle n_{\rm H} \rangle  = 4.4 \times 10^{-5}$ cm$^{-3}$, and the 
total gaseous mass, including helium, is
\begin{equation}
   M_{\rm gas} = (4 \times 10^{11}~M_{\odot}) 
       \left[ \frac {R}{500~{\rm kpc}} \right]^2
       \left[ \frac {\rm N(H~I)} {2 \times 10^{16}~{\rm cm}^{-2}} 
       \right] ^{1/2} \; .
\end{equation}
The gas mass could therefore approach 1\% of the $10^{14}~M_{\odot}$ 
required to bind the group, a relatively small gas 
fraction compared to the gas found in clusters
and elliptical-rich groups.  Likewise, the luminous mass in these 
galaxies is small.  If the gas is clumped into denser parcels
with the same total covering factor, the neutral fraction would increase 
and the total gas mass would decrease inversely with the characteristic 
depth of the absorbers.   

Let us now consider inhomogeneous models for the gas distribution around
\PKS, involving dwarf galaxies or primordial gas filaments.
Let the total projected area of the gas be $L \times L$ with depth $D$,
where $L \approx 1$ Mpc and $D \approx 400$ kpc. Assume that this volume
is filled by an ensemble of $N_{\rm cl}$ clouds, each of radius 
$r \approx 40$ kpc, with a velocity dispersion 750 \kms.    
The cloud filling factors in area and volume are, 
$f_A = \pi N_{\rm cl} (r/L)^2$ and $f_V = (4 \pi/3) N_{\rm cl} (r/L)^3 (L/D)$, 
where $f_A$ could exceed 1 if clouds overlapped in projected area. 
Using the given parameters, their ratio is $(f_A/f_V) = (4/3)(r/D) 
\approx 0.13$.  Since we detected many absorbers along the  
sightline, it is likely that $f_A \geq 1$, which suggests that 
$f_V \approx 0.3-0.5$.  A consequence of this model is that 
these ``clouds'' should collide with one other 
on a crossing time $t_{\rm cr} $= (0.6 Mpc)/(750 \kms) $< 10^9$ yr.   
Therefore, if this group is bound, we would expect considerable gas 
stripping, with the possibility of shocks, hot gas 
($T \approx 5 \times 10^6$~K), and tidal plumes.  There is no
evidence for these effects, although there has been no search for
extended soft X-ray emission or O~VI absorption from this region.  

    If the group of galaxies and clouds is not bound, then this
cloud ensemble could either be primordial gas which never participated
in galaxy formation or gaseous halos of luminous or dwarf galaxies.
Based upon an extensive survey of quasar fields, for which ultraviolet
spectra were obtained by the HST Key Project Team, 
Lanzetta et al. (1995, hereafter L95) identified $\sim1/3$ of large
equivalent width ($W_{\lambda} \geq 0.3$ \AA) \Lya\ absorption lines with
bright galaxies at impact parameters $\leq 210 h_{75}^{-1}$ kpc away from 
the sightline. The three strongest close pairs of \Lya\ clouds in Table
1 would not be resolved into multiple components at the 1 \AA\ 
resolution of the Key Project spectra.  While all three 
cloud complexes are easily strong enough
to be included in the L95 study, none is 
as close to a bright galaxy as those in the L95 survey: 
$300 h_{75}^{-1}$ kpc for the 5100 \kms\ absorber, 
$560 h_{75}^{-1}$ kpc for the 16,300 \kms\ absorber, and 
$400 h_{75}^{-1}$ kpc for the 17,000 \kms\ absorber. Our
H~I detection survey in this field (van Gorkom et al. 1996 and \S~2.2 
herein) reaches depths at least comparable to the L95 survey (0.1--0.3 $L_*$). 
Thus, the bright galaxy halos in this field may be anomalously larger
than those found by L95 in other fields, and also substantially larger
than expected theoretically (Dove \& Shull 1994).  
However, we strongly suspect that bright galaxy halos cannot 
account for these clouds. 

Are there undetected dwarf galaxies at the \Lya\ cloud redshifts 
closer to the \PKS\ sightline?  If so, then dwarf-galaxy halos could
be responsible for these absorptions as extrapolated by L95 and others.
Chen \etal (1998) found a best-fit \Lya\ halo size that scales with
galaxy luminosity as $r_h \propto  L^{0.35}$, a somewhat stronger 
dependence than the $r_h \propto L^{0.15}$ found for Mg~II absorbing 
halos (Steidel 1995). With the Chen \etal scaling relation, 
the two weaker absorption complexes (5100 and 16,200 \kms) would be
consistent with the L95 and Chen \etal observations if any 
fainter galaxies ($L \leq 0.1 L_*$) were found closer to \PKS\ at those 
redshifts.  However, the 17,000 \kms\ absorber is so strong that a dwarf 
galaxy would need to be very close to the sightline to satisfy the 
Chen et al.  halo-size relationship.  Specifically, we would 
require impact parameters $\rho < 67 h_{75}^{-1}$ kpc for 
$L \approx 0.1 L_*$ and $\rho < 33 h_{75}^{-1}$ kpc for 
$L \approx 0.01 L_*$. The only bright galaxy sufficiently 
near to \PKS\ to satisfy the first condition is galaxy
G4 of Falomo \etal (1993), which has $L \approx 0.2 L_*$
and $\rho\approx 80 h_{75}^{-1}$ kpc. However,  G4 has $z = 0.117$, the 
same as \PKS.  While there are some closer galaxies, their numbers 
are modest:  1 -- 3 excess galaxies to $L \leq 0.01 L_*$ based on the 
independent data from Wurtz et al. (1997) and Falomo \etal
(1993).  The only two galaxies with spectroscopy in hand 
are galaxies 1 and 2 of Falomo \etal (1993) at $z = 0.117$, located
4 arcsec E and 25 arcsec SE of the BL Lac, respectively.  At this
point, there is no dwarf galaxy candidate for any of these three strong
absorption complexes.  

Thus, although unlikely, the dwarf-galaxy halo
hypothesis remains possible.  We detected one faint dwarf
($M_B \approx -15$) at $565 h_{75}^{-1}$ from the \PKS\ sightline,  
and we may find other faint galaxies at the redshifts of the three 
strong absorption complexes.  However, there is no direct evidence 
in favor of the dwarf-halo, and the primordial cloud hypothesis must 
be taken seriously for all three \Lya\ absorbers.

\subsection*{3.2. Metallicity Limits }

We can also use our spectra to set limits on the metallicity of the 
strongest \Lya\ absorbers.  As seen in Figure 1, metal lines should be 
searched for in three strong absorption systems,
corresponding to \Lya\ lines at $\lambda_1 = 1281.393$~\AA\ ($z_1 = 
0.054063$), $\lambda_2 = 1284.484$~\AA\ ($z_2 = 0.056605$), 
and $\lambda_3 = 1285.097$~\AA\ ($z_3 = 0.057110$).
In the limited wavelength range of our GHRS/G160M spectrum, the only 
expected strong metal line is the Si~III $\lambda1206.500$ resonance line. 
No Si~III absorption is present at the 1271.73~\AA\ location of
system 1.   Although we see a hint of Si~III absorption
at the expected positions [1274.79 \AA\ and 1275.40 \AA,
see Fig. 1] corresponding to systems 2 and 3, we treat this absorption
as {\it upper limits}.  Using the unsmoothed data to detect weak,
unresolved features, we obtain a formal $4 \sigma$ error on equivalent 
width of 23 m\AA\ (rest-frame 22 m\AA).

To convert the observed Si~III/H~I to limits on abundances (Si/H), 
we must make an ionization correction based on photoionization conditions 
in the absorbers. Appenzeller et al. (1995) claimed to detect weak
Lyc absorption with the far-UV spectrograph aboard {\it ORFEUS}.
The \Lya\ clouds at $z = 0.054-0.057$ were estimated to have a combined 
column density N(H~I) = $(2-5) \times 10^{16}$ cm$^{-2}$. We have 
derived the expected column 
densities of metal ions by modeling the strong absorbers as slabs with
N(H~I) = $2 \times 10^{16}$ cm$^{-2}$ and total depth ranging
from 5 to 800 kpc, comparable to or less than the offset distance from 
the nearest galaxies.  The slabs are assumed to be illuminated on both 
sides by an ionizing spectrum with specific intensity at 1 Ryd of 
$I_0 = 10^{-23}$ ergs~cm$^{-2}$ s$^{-1}$ Hz$^{-1}$ sr$^{-1}$,
consistent with a recent calculation by Shull \etal\ (1998) that
gives $I_0 = (1.1 \pm 0.5) \times 10^{-23}$, based on low-$z$ Seyfert galaxies 
and a new IGM opacity model.  The spectral index of the
ionizing background is taken as $\alpha_s = 1.8 \pm 0.1$ (Zheng \etal 1997). 

As a ``standard model'', we assume a homogeneous slab, of depth 400 kpc 
and 0.003 solar metallicity.  The photoionization equilibrium
is computed with the model CLOUDY Version 90.03 (Ferland 1996). 
The results are 
given in Table 4. In general, the C~III $\lambda977$ and C~IV $\lambda1548$ 
lines are the best tracers of metals, while Si~III $\lambda1206$ is a 
factor of 5--10 weaker, assuming relative solar abundances of 
(Si/H)$_{\odot} = 3.55 \times 10^{-5}$ and 
(C/H)$_{\odot} = 3.63 \times 10^{-4}$ (Grevesse \& Anders 1989). 
However, the C~III line lies shortward of the HST band,
and its observation must await the launch of the FUSE satellite in early
1999.  For unsaturated absorption, the predicted equivalent widths can be 
written:
\begin{eqnarray}
   W_{\lambda}(\rm C~IV) = (73~{\rm m\AA}) 
      \left[ \frac {\rm N(H~I)} {2 \times 10^{16}~{\rm cm}^{-2}} \right]  
      \left[ \frac {\rm [C/H]} {0.003} \right] \; , \\
                \nonumber \\
   W_{\lambda}(\rm Si~III) = (22~{\rm m\AA})
      \left[ \frac {\rm N(H~I)} {2 \times 10^{16}~{\rm cm}^{-2}} \right]
      \left[ \frac {\rm [Si/H]} {0.003} \right] \; . 
\end{eqnarray}
Following the convention of our earlier \Lya\ work, we treat
``definite absorbers'' as those with $4 \sigma$ significance. 
The observed 23 m\AA\ ($4 \sigma$) limit on Si~III $\lambda1206$ 
(rest equivalent width 22 m\AA) corresponds to N(Si~III) 
$\leq 1.0 \times 10^{12}$ cm$^{-2}$.  If we scale the column density 
of the strong absorption components to N(H~I)$ = 2 \times 10^{16}$ cm$^{-2}$,
the ($4 \sigma$) upper limit on metallicity can be written,
\begin{equation}
   \left[ \frac {\rm Si} {\rm H} \right] \leq (0.003) 
          \left[ \frac {2 \times 10^{16}~{\rm cm}^{-2}} {\rm N(H~I)} \right]
          \left( \frac {\rm Si} {\rm H} \right)_{\odot} \; . 
\end{equation}
 
\begin{figure}
   \epsscale{0.85}
   \plotone{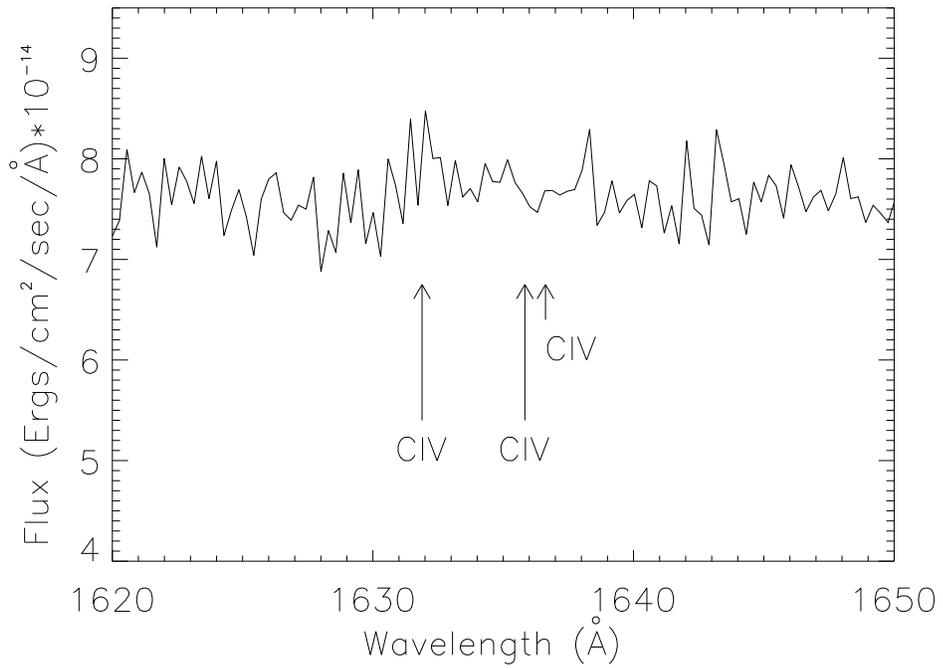} 
   \caption[]{GHRS/G140L data of \PKS\ (Bruhweiler \etal 1993) 
    reduced with our software and GHRS final calibration files 
    to show the spectral region for C~IV 
    $\lambda1548.195$ absorption corresponding to the three strong 
    \Lya\ absorbers at 1281.393, 1284.484, and 1285.097 \AA\ (see Table 4). 
    The signal-to-noise ratio is 30 per 0.029 \AA\ pixel.
    The C~IV upper limits are each 134 m\AA\ ($4\sigma$) in the 
    observed frame. } 
\end{figure}

Another limit on metallicity comes from C~IV $\lambda1548.195$.
From their low-resolution (GHRS/G140L) spectrum,
Bruhweiler \etal (1993) quote a ($2 \sigma$) upper limit of 110 m\AA\ for
C~IV $\lambda1548$. We recalibrated the pre-COSTAR GHRS/G140L spectrum using
IRAF/STSDAS/CALHRS with the final calibration files. As before, 
subexposure coaddition was performed using our own IDL routines,
which merge the subexposures by exposure time with photocathode
blemishes removed (Fig. 5).  We find a $4 \sigma$ limit of 
$W_{\lambda} \leq 134$ m\AA\ (127 m\AA\ rest frame) or
N(C~IV) $\leq 3 \times 10^{13}$ cm$^{-2}$ for a linear curve of
growth. This yields a metallicity limit,
\begin{equation}
   \left[ \frac {\rm C} {H} \right] \leq (0.005)
          \left[ \frac {2 \times 10^{16}~{\rm cm}^{-2}} {\rm N(H~I)} \right]
          \left( \frac {\rm C} {\rm H} \right)_{\odot} \; .
\end{equation}

\begin{figure}
  \epsscale{0.85}
  \plotone{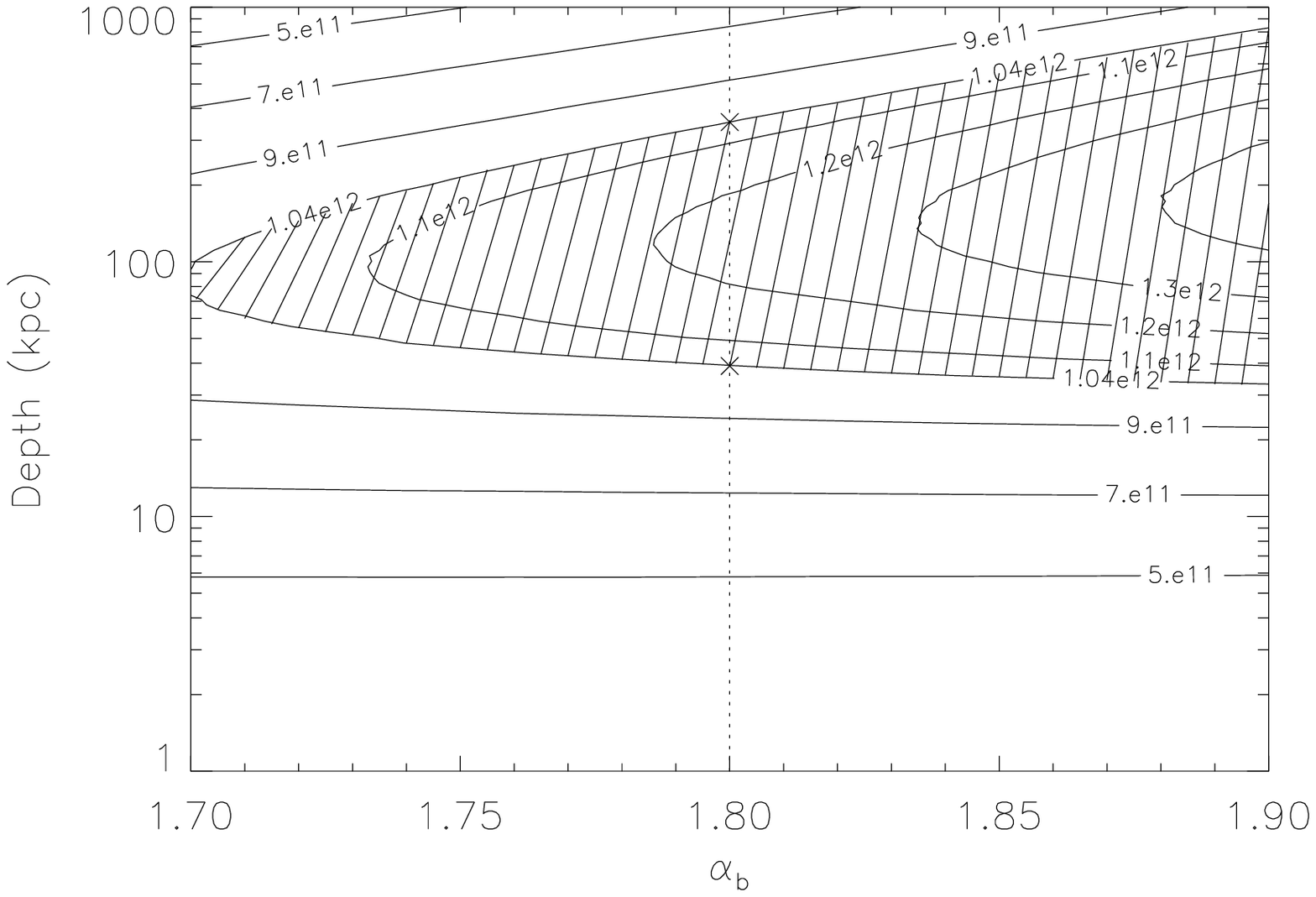}
   \caption[]{ Contours of column density, N(Si~III), 
   corresponding to N(H~I) $= 2 \times 10^{16}$ cm$^{-2}$ and 0.003 
   solar metallicity. The axes represent the assumed spectral slope
   of the ionizing background spectra ($1.7 \leq \alpha_b \leq 1.9$)
   and cloud depths ($5 \leq D \leq 1000$ kpc). Hatched lines
   mark regions disallowed by the $4 \sigma$ non-detection of Si~III,  
   while asterisks mark the first viable solutions, $D \geq 300$ kpc
   and $D \leq 40$ kpc, consistent with the standard value 
   $\alpha_b = 1.8$.  Large-$D$ solutions correspond to 
   absorption from a homogeneous slab, while small-$D$  
   solutions could arise in denser galactic halos along the sightline. } 
\end{figure}

\begin{figure}
  \epsscale{0.85}
  \plotone{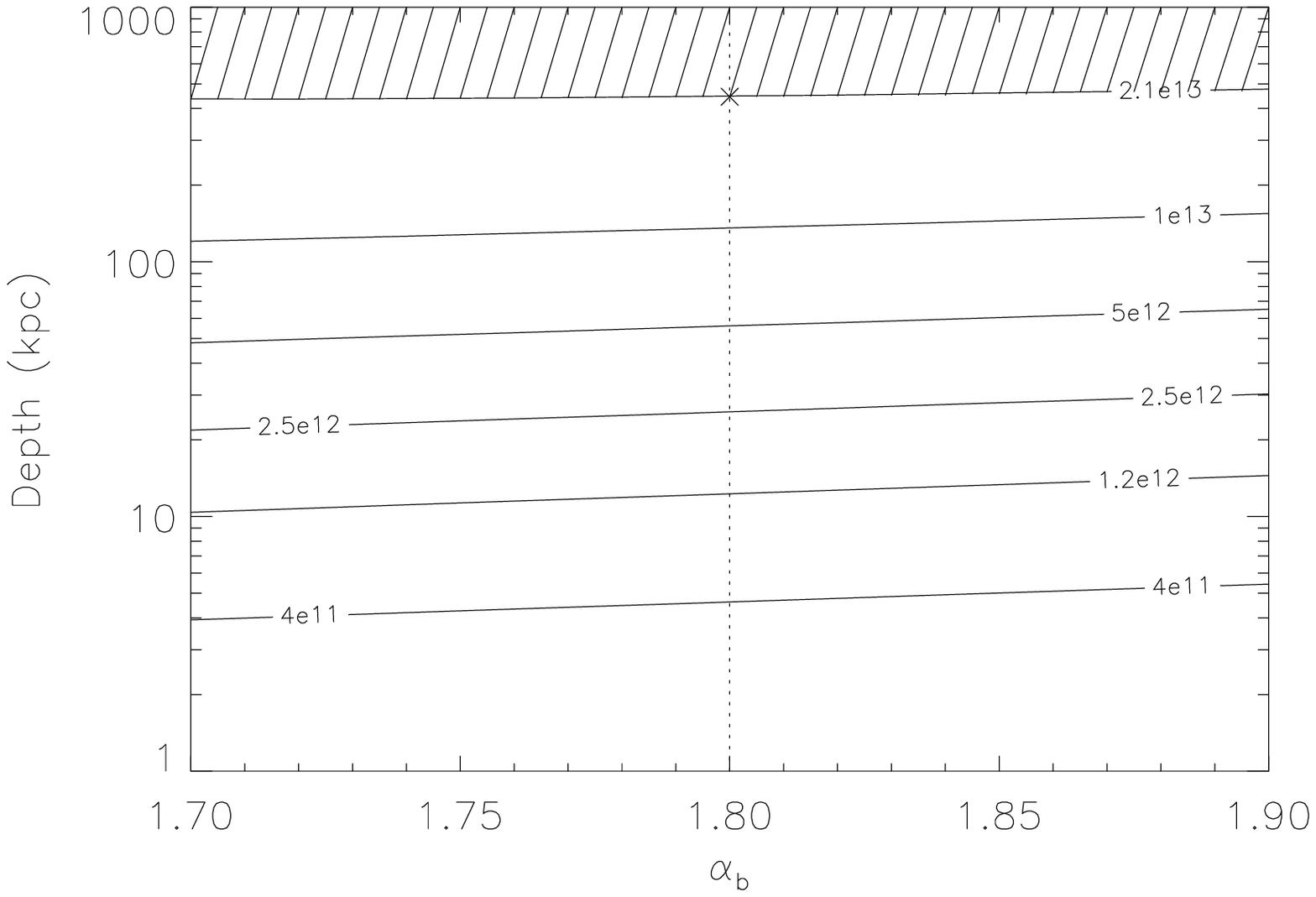}
   \caption[]{ Same as Fig. 6, showing contours of column density, 
    N(C~IV), corresponding to N(H~I) $= 2 \times 10^{16}$ cm$^{-2}$ 
    and 0.003 solar metallicity. Hatched lines mark regions disallowed 
    by the $4 \sigma$ non-detection of C~IV, while the asterisk marks 
    the first viable solutions, $D \leq 400$ kpc, consistent with 
    $\alpha_b = 1.8$. Although the combined Si~III and C~IV limits
    allow metal-bearing clouds of size 200--400 kpc, there is more
    parameter space at sizes less than 40 kpc. }
\end{figure}

Figures 6 and 7 illustrate how these metallicities depend on the two key 
parameters of the photoionization models: the mean hydrogen density 
$\langle n_H \rangle$ and the spectral 
slope $\alpha_b$ of the ionizing background. The observed column density,
N(H~I), and the assumed cloud depth, $D$, determine the mean neutral density, 
\begin{equation}
    \langle n_{\rm HI} \rangle = (1.6 \times 10^{-8}~{\rm cm}^{-3})
     \left[ \frac {\rm N(H~I)} {2 \times 10^{16}~{\rm cm}^{-2}} \right]
     \left[ \frac {400~{\rm kpc}} {D} \right]   \; .
\end{equation}
The neutral fraction, $\langle n_{\rm HI} / n_H \rangle$, is set by 
photoionization equilibrium and is proportional to the
ionization parameter, $U \propto J_0/n_H$, as are the other metal-ion 
column densities.  The precise behavior of the ratios (Si~III/H~I) and
(C~IV/H~I) depend on the range of $U$ where these particular ions
peak in fractional abundance (Donahue \& Shull 1991).   
The spectral slope, $\alpha_b$, affects the 
ionization rates above the relevant ionization thresholds (16.34, 33.46,
45.13 eV for Si~II, III, IV; 24.38, 47.87, 64.48 eV for C~II, III, IV).  
For a fixed radiation intensity, $J_0$, decreasing the assumed cloud depth 
$D$ results in a higher $\langle n_H \rangle$, smaller ionization parameter
$U$, larger hydrogen neutral fraction, and lower N(C~IV) and N(Si~III).
For $\alpha_b = 1.8 \pm 0.1$ and $D = 400 \pm 200$ kpc, the predicted
column densities are:
\begin{eqnarray}
 {\rm N(Si~III)} &=& (1.0^{+0.2}_{-0.2} \times 10^{12}~{\rm cm}^{-2})
               \left( \frac {\rm [Si/H]} {0.003} \right)  \\ 
            \nonumber      \\
 {\rm N(C~IV)}   &=& (1.8^{+0.7}_{-0.5} \times 10^{13}~{\rm cm}^{-2})
          \left( \frac {\rm [C/H]} {0.003} \right)  \; .   
\end{eqnarray}

The topology of C~IV and Si~III differ, as seen in the excluded
regions of Figures 6 and 7.  For denser absorbers, the ionization
equilibrium shifts to lower ionization parameter ($U$) in which
case C~IV/H~I decreases (Fig. 7) while Si~III/H~I remains about the same
(Fig. 5).  For Si~III, the ionization solutions can be double-valued,
since Si~III changes more strongly with $U$ than H~I (Donahue \&
Shull 1991).  At sufficiently low and sufficiently high
ionization parameter, silicon lies in other states (Si~II or Si~IV).
As a result, the H~I absorption toward \PKS\ could still contain
metals at 0.01 solar and arising either 
in a homogeneous slab, of depth $D = $200-400 kpc, or in denser parcels 
such as 10--40 kpc halos of dwarf galaxies or intergalactic clouds.  

Thus, it is easier to hide metals from detection in the C~IV lines than 
in Si~III.  The observed limit on Si~III absorption with N(H~I) = 
$2 \times 10^{16}$ cm$^{-2}$ and 0.003 metallicity formally allows 
solutions with $D > 400$ kpc or $D < 40$ kpc.  Limits 
on C~IV absorption allow $D < 400$ kpc.  The combined C~IV and Si~III
limits, together with the numerous discrete \Lya\ absorbers
in velocity space, suggest that the gas may be inhomogeneous 
on scales of tens of kpc.   However, it is puzzling 
how clumped absorbers could maintain a high density contrast
when these clumps would be expected to collide on a billion-year time 
scale. Such collisions would also shock-heat the gas
possibly making C~IV lines more detectable.   
Filamentary sheets, with large aspect ratios, provide
a more plausible scenario consistent with these constraints.  

Before concluding this section, it is worth examining the accuracy
of the measurements of N(H~I).  Our assumed value,
N(H~I) $= 2 \times 10^{16}$ cm$^{-2}$, comes from the
claimed detection of a Lyc depression in {\it ORFEUS} data
(Appenzeller \etal 1995).  These data suggest a flux
discontinuity shortward of 970 \AA, corresponding to the $z =  
0.054 - 0.057$ \Lya\ absorbers, but the Lyman break
is not definitive, owing to the moderate (0.5 \AA) 
spectral resolution, low signal-to-noise
(S/N = 15), and uncertainties in defining and extrapolating the true 
continuum between 920 and 1100 \AA.   For similar reasons, these data 
are not sufficiently accurate to determine N(H~I) from a curve-of-growth
analysis of the higher Lyman series, Ly$\beta$ -- Ly$\delta$. 

Because of \Lya\ line saturation in our HST/GHRS data, 
we cannot confirm the large H~I columns implied by the Lyc 
absorption.   However, we can provide both a firm lower limit 
and give reasonable estimates from the strongest \Lya\ features at 
1284--1285 \AA.  The optical depth of the \Lya\ line per unit velocity
is $\tau(v) = (\pi e^2/m_ec) f \lambda N(v)$, where $N(v)$ is the H~I
column density per unit velocity. We may integrate this, for
$f = 0.4164$ and $\lambda = 1215.67$ \AA, to obtain, 
\begin{equation}
   N_{\rm tot} = (7.45 \times 10^{11}~{\rm cm}^{-2}) \int \tau(v) \; dv \;, 
\end{equation}
where $v$ is measured in km~s$^{-1}$.  In our post-COSTAR data, 
with S/N $\approx 20$, this formula provides only a minimum 
column density, since we cannot distinguish optical depths $\tau(v) > 3$.
Flux calibration and background subtraction are difficult with
the GHRS one-dimensional detectors. 
An integration of the data (Fig. 8) yields values 
$N_{\rm tot}$ ranging from $(3 - 10) \times 10^{14}$ cm$^{-2}$.  
Values of $2 \times 10^{16}$ cm$^{-2}$ needed for consistency with the
{\it ORFEUS} Lyc measurement would require unresolved narrow components
in the line core.   

Although we have some skepticism about adopting the 
($2 \times 10^{16}$ cm$^{-2}$) column density from {\it ORFEUS},
it is worth remembering that line saturation and multiple velocity 
components make column density measurement extremely difficult
from just one line.  A prime example of this
difficulty comes from the \Lya\ absorbers seen toward 3C~273 at 
$cz \sim 1000$ \kms\ and $cz \sim1600$ \kms\ (Weymann et al. 1995).   
Recent {\it ORFEUS} measurements of these components 
(Hurwitz et al. 1998) find Ly$\beta$ absorption
equivalent widths larger by factors of 1.5 and 2.4
than values predicted by \Lya\ Voigt profile fits of HST/GHRS data 
($\log N = 14.19$ and $b = 40.7$ \kms\ for the 1000 \kms\ cloud, and 
$\log N = 14.22$, $b = 34.2$ \kms\ for the 1600 \kms\ cloud).  
By analyzing both \Lya\ and Ly$\beta$,  
Hurwitz et al. (1998) derive a best-fit of $\log N = 15.8$ and
$b = 17$ \kms.  They conclude that, if the H~I absorption arises in a 
single cloud, its column density is higher by at least a factor of 4 
compared to the value of Weymann et al. (1995).

\begin{figure}
  \plotone{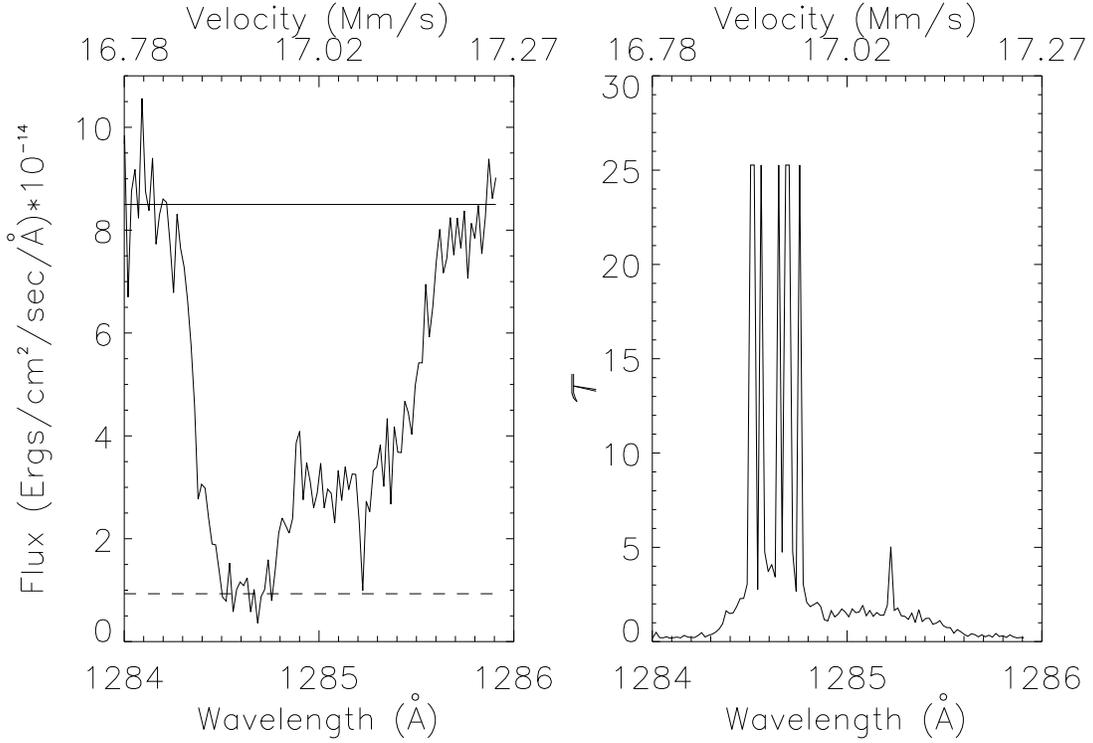}
   \caption[]{Blow-up of the \Lya\ absorption cluster 
   at $cz \approx 17,000$ \kms.  
   Left:  \Lya\ absorption line vs. wavelength and heliocentric 
   velocity, $cz$, showing the continuum flux (solid line at top) and 
   $1 \sigma$ background (dashed line at bottom).  
   Right: optical depth, $\tau(v)$ of the same absorption line, computed 
   from left panel, assuming that the continuum lies at the dashed line. 
   We ignore negative fluxes and values $\tau > 25$. Owing to 
   uncertainties in flux calibration and background subtraction,
   values $\tau > 3$ are unreliable.  The integral of this absorption 
   yields N(H~I) =  $(3-10) \times 10^{14}$ cm$^{-2}$ (the higher value
   for right panel shown here).  Larger columns, consistent with the
   {\it ORFEUS} LyC estimates, N(H~I) $=(2-5) \times 10^{16}$ cm$^{-2}$,
    would require unresolved narrow components in the line core. } 
\end{figure}

Thus, it could be that the strong, saturated \Lya\ lines  
towards \PKS\ have columns well above $10^{15}$ cm$^{-2}$
and even as high as $2 \times 10^{16}$ cm$^{-2}$.   
This could arise if narrow H~I components 
are hidden within the line core at $17,000 \pm 50$ \kms.  
For example, 20 components, each with $b \approx 15$ \kms\
and N(H~I) $\approx 10^{15}$ cm$^{-2}$, spread stochastically
over $\sim100$ \kms\  would reproduce the \Lya\ and Lyc data. 
Each component would have central optical depth $\tau_0 \approx 50$
and an (isolated) equivalent width of 240 m\AA.  Saturation
and velocity overlap would allow their accumulated \Lya\ absorption to  
give the observed $\sim0.8$ \AA, while the Lyc optical depth could
reach values $\tau_c \approx 0.1$.      

Without better data on the Lyc absorption edges or
the higher Lyman series lines, we cannot verify the \Lya\ absorption
($2 \times 10^{16}$ cm$^{-2}$) in the strong absorber at 
$cz = 16,970$ km~s$^{-1}$.  Confirmation of the Lyman limit and 
higher Lyman series will await our planned studies of \PKS\ 
following the scheduled Feb. 1999 launch of the Far Ultraviolet 
Spectroscopic Explorer (FUSE).  
With FUSE, if N(H~I) is as large as $2 \times 10^{16}$
cm$^{-2}$, the Lyman edge at 963.36~\AA\ will have $\tau_c \approx 0.13$, 
and many higher Lyman lines will be detectable.   
Even if we adopt the minimum values,
N(H~I) $\geq 4 \times 10^{14}$ cm$^{-2}$ and $\tau({\rm Ly}\alpha) \geq 3$ 
in the line core, we should be able to detect higher Lyman lines
with $\tau({\rm Ly}\beta) \geq 0.48$, $\tau({\rm Ly}\gamma) \geq 0.17$, 
and $\tau({\rm Ly}\delta) \geq 0.08$.  It would also be helpful to 
obtain a better \Lya\ spectrum of the 1270-1290
\AA\ region with HST/STIS, whose two-dimensional array detectors
will improve the background subtraction and flux calibration.
The metal-line searches can also 
be improved with HST, using STIS and COS (see \S~3.4).   

If N(H~I) is smaller than the value ($2 \times 10^{16}$ cm$^{-2}$) 
assumed in Figures 5 and 6, our limits on metallicity increase.  
For homogeneous clouds, of constant density, several quantities follow 
from simple scaling relations of N(H~I).  First, our inferred
value of D/H scales inversely with N(H~I), and would therefore
go up.  Second, one would find a 
tradeoff of cloud depth ($D$) and metallicity ([C/H] or [Si/H]).
If N(H~I) were decreased by a factor 10, the metallicity limits
would rise by the same factor. If $D$ is held constant, a simple scaling
of the metallicity with N(H~I) is not possible.
As we discussed earlier, the relative ionization fractions
of H~I, C~IV, and Si~III all vary with ionization parameter, $U$.  
If we retain our estimate of the radiation background and
assume $D = 400$ kpc, reducing N(H~I) by a factor of 10 changes 
the metallicity limits to C/H$< 0.023$ and Si/H$< 0.14$ times solar.
For the range in ionization parameter
$-1.9 < \log U < -1.3$ appropriate to these
assumptions about cloud density and radiation field,
the greater sensitivity to $U$ of the Si~III fraction
causes the Si~III constraints on metallicity to
weaken much more rapidly with decreasing N(H~I).
Regardless of the exact value of N(H~I), the most sensitive metal
lines remain upper limits.  Therefore, one must take seiously 
the possibility that these clouds are primordial.  
 
\subsection*{3.3. Limits on Deuterium }

A similar analysis can be done using the strong absorbers to search for 
weak deuterium \Lya\ (1215.339~\AA) features in the shortward wings of 
H~I \Lya\ (1215.667 \AA).  No deuterium feature is seen in the wings of 
the H~I system 1 (H~I at 1281.393~\AA, D~I at 1281.044~\AA) 
to a level of $W_{\lambda} \leq 20$~m\AA.  In 
system 2 (H~I at 1284.484~\AA, D~I at 1284.133~\AA)
the limit is $W_{\lambda} \leq 30$~m\AA.  For a 
linear curve of growth, these equivalent widths translate to
N(D~I) $\leq 3.7 \times 10^{12}$ cm$^{-2}$ and 
$5.5 \times 10^{12}$ cm$^{-2}$, respectively.  If we attribute column 
densities N(H~I) of $1 \times 10^{16}$ cm$^{-2}$ and 
$2 \times 10^{16}$ cm$^{-2}$  to systems 1 and 2, respectively, these 
limits give a deuterium abundance limits of,
\begin{eqnarray}
    \left( \frac {\rm D}{\rm H} \right) \leq 
    (3.7 \times 10^{-4}) \left[ 
    \frac {1 \times 10^{16}~{\rm cm}^{-2}} {\rm N(H~I)} \right] 
         \; \; {\rm (system~1)} \\ 
                                              \nonumber    \\ 
 \left( \frac {\rm D}{\rm H} \right) \leq 
    (2.8 \times 10^{-4}) \left[ 
    \frac {2 \times 10^{16}~{\rm cm}^{-2}} {\rm N(H~I)} \right] 
         \; \; {\rm (system~2)}  
\end{eqnarray}

These limits can be improved significantly with higher-precision 
data on the D~I line (with HST/STIS) and with better H~I column densities
from higher Lyman-series lines (with the FUSE spectrograph).
At the current level of accuracy, neither limit provides a strong constraint 
on Big Bang nucleosynthesis (Schramm \& Turner 1998).  The interstellar medium
value has been reported in the range D/H $= (1.6 \pm 0.09) \times
10^{-5}$ (Linsky et al. 1993, 1995).   
However, a controversy still exists over the high-redshift D/H observed
in QSO absorption systems. A recent determination
of D/H in two QSO Lyman-limit systems (Burles \& Tytler 1997) 
gives a ``low value'',  D/H $= (3.4 \pm 0.25) \times 10^{-5}$, while 
Songaila and Cowie (cf. Songaila 1997) quote ``high values'' with a range 
D/H $= (0.4  - 1.5) \times 10^{-4}$.  The high values of D/H would
obviously require a large destruction rate of deuterium through
star formation.  Thus, it would be helpful to obtain a detection of 
deuterium in the strong \Lya\ systems at $z = 0.054-0.057$ toward \PKS.
Here, the measured metallicity is much less than solar values,
implying little astration of D.  We expect that 
future HST and FUSE observations can make a factor of 3 improvement in
sensitivity, which could set a limit on D/H $<1 \times 10^{-4}$.

\subsection*{3.4. Future Searches for Metal Absorption} 

Our photoionization models demonstrate that, for clouds with 
$\geq0.001$ solar metallicity, it should
be possible, with better HST/STIS data, to detect the 1548.2 \AA\ and 1550.8
\AA\ resonance lines of C~IV.  We may also be able to detect C~II
$\lambda1335$,  N~V $\lambda1238$, Si~IV $\lambda1394$, and 
Si~III $\lambda1206$.  With the FUSE satellite, we intend to search for
C~III $\lambda977$, C~II $\lambda1036$, and O~VI $\lambda1032$.  The C~III
line should be detectable down to well below 0.001 solar metallicity.
The C~II lines might be present if the gas is at higher density
(lower $U$).  The O~VI line is weaker in the standard model 
($D = 400$ kpc, $\alpha_b = 1.8$), but it might be enhanced by
a number of effects:  a hard photoionizing radiation field above 114 eV, 
a collisionally ionized component due to hot gas,
or oxygen abundance enhancement by massive-star nucleosynthesis.
We probably will not detect O~I $\lambda1302$, which is weak, even at
3\% solar metallicity, because the O~I abundance is tied by charge
exchange to the ratios (H~II/H~I) and (O~II/O).  

Because of the O~VI ionization correction, the strength of
the $\lambda 1032$ absorption line depends both
on the shape of the ionizing spectrum and on the cloud depth.
For example, if $\alpha_b = 1.4$ ($D = 400$ kpc) the O~VI column density
increases by over a factor of 4, while if $\alpha_b = 1.8$ ($D = 600$ kpc)
it increases by a factor of 2.  The {\it total} H~I column in 
the absorbers could be higher than the assumed value, $2 \times 10^{16}$ 
cm$^{-2}$.  In addition, the Si and O abundances could be enhanced 
(Songaila \& Cowie 1996; Giroux \& Shull 1997) relative to C in regions 
dominated by ``prompt nucleosynthesis'' from massive-star supernovae. Thus, 
if Si/C $>2 \times ({\rm Si/C})_{\odot}$ and 
O/C $>2 \times ({\rm O/C})_{\odot}$, 
we may be able to detect lines of O~VI, Si~II, Si~III, and Si~IV.  

As shown by Donahue \& Shull (1991), 
the ratios of column densities from multiple ion states of the same element 
(e.g., Si~II/III/IV and C~II/IV) can be modeled to provide diagnostics of 
the intensity and spectrum of the 1--4 Ryd ionizing radiation field. 
Detections of these ions would allow us to discriminate between an 
extragalactic radiation field due to AGN or due to
O-stars from starburst galaxies.

\section*{4. CONCLUSIONS }

The sightline to \PKS\ is unique among the AGN studied
thus far for low-redshift \Lya\ absorbers.   It has the highest
frequency of absorbers, and they are identified with a  
concentration of bright galaxies.  The \Lya\ absorbers
at $cz \approx 17,000$ \kms\ may be an extreme example of the
previously known association of \Lya\ clouds with the extended
halos of galaxies (Lanzetta \etal 1995; Stocke \etal 1995).  
Of greater interest is the enormous inferred gaseous extent;
the nearest bright galaxies lie $(400-800)h_{75}^{-1}$ off the sightline.   
These offsets are so large that they are unlikely to represent
equilibrium gaseous disks;  the orbital times at such distances
would be enormous.  We have interpreted these absorbers as large 
sheets of intragroup gas, or as smaller primordial clouds 
and halos of dwarf galaxies.  The clumpiness of the \Lya\ absorption
in velocity space suggests that some spatial
structure is present.  However, as  discussed in \S~3.1,
a medium that is clumpy in both space and velocity would be
subject to disruptive collisions.  

Our major conclusions are therefore:  
\begin{itemize}

\item The metallicity limits for the \Lya\ absorbers give (Si/H) and
(C/H) less than 0.003 solar if N(H~I) = $2 \times
10^{16}$ cm$^{-2}$ (the value from ORFEUS). The greatest uncertainty
in these metallicities is the precise value of N(H~I). 
We will use the FUSE satellite to measure the H~I columns 
through Lyman limit absorption and higher Lyman series lines. 
We hope to see discrete absorption edges at 
961.04, 963.36, and 963.82 \AA.  

\item  Our metallicity limit, [Si/H] $<0.003$ solar contradicts the
hypothesis that these clouds are metal-enriched intragroup gas.  
Mushotzky \& Loewenstein (1997) suggest that 
intracluster gas might be enriched early (at $z > 0.4$) to levels of 
10\% solar metallicity.  Evidently, the intergalactic gas around \PKS\ 
has not been enriched to the levels observed in X-ray emitting
intracluster gas.

\item If the ``small group'' hypothesis is correct, a search for X-rays
with AXAF imagers would test whether any hot gas has been produced
by tidal effects or stripping.   

\item Future spectral observations with HST can provide even better
limits on D/H, Si~III, and C~IV than those reported here.  With FUSE,
we will be able to search for the expected strong C~III
$\lambda 977$ line, the weak O~VI $\lambda1032$,
as well as absorption lines of S~II, C~II, S~II, and
N~II which may  be present if the absorption occurs in denser halos.
 
\item The low metallicity limits and the large projected distances
from the nearest galaxies suggest that these clouds could be primordial.

\end{itemize}

\vspace{1cm}

This work was based on observations with the NASA/ESA {\it Hubble Space
Telescope} obtained at the Space Telescope Science Institute,
which is operated by AURA, Inc. under NASA contract NAS5-26555
and on observations made with NRAO's Very Large Array. The NRAO
is operated by Associated Universities, Inc. under a cooperative
agreement with the National Science Foundation.
We thank Michael Fall and Richard Mushotzky for helpful discussions
on the evolution of intracluster gas.  This work was supported by HST Guest
Observer grant GO-06593.01-95A and by the Astrophysical Theory Program
(NASA grant NAGW-766 and NSF grant AST96-17073) to the University of Colorado
and by an NSF grant (AST96-17177) to Columbia University.

\newpage

\begin{center}
{\bf Table 1} \\
{\bf HST Absorption Features$^a$ in \PKS } \\
\   \\
\begin{tabular}{ccccc}
\hline\hline 
Wavelength         & Velocity$^b$    & Rest~EW      & Significance$^c$ & ID \\
 (\AA)             &  (\kms)         &   (m\AA)     & ($\sigma$)       & \\
\hline
$1226.362\pm0.054$ & $2637\pm13$     & $42\pm24$    & 9.0  & \Lya  \\
$1226.990\pm0.062$ & $2789\pm15$     & $36\pm18$    & 8.0  & \Lya  \\
$1232.032\pm0.045$ & $4035\pm11$     & $21\pm11$    & 4.2  & \Lya   \\
$1235.941\pm0.050$ & $4999\pm12$     & $129\pm7$    & 29   & \Lya   \\
$1236.436\pm0.100$ & $5121\pm24$     & $201\pm8$    & 45   & \Lya   \\
$1238.426\pm0.099$ & $5612\pm24$     & $21\pm10$    & 4.8  & \Lya   \\
$1238.744\pm0.065$ & $5690\pm16$     & $85\pm22$    & 20   & \Lya   \\
$1239.817\pm0.040$ &  Galactic       & $33\pm12$    & 7.5  & Mg~II \\
$1240.411\pm0.052$ &  Galactic       & $20\pm11$    & 4.5  & Mg~II \\
$1250.588\pm0.015$ &  Galactic       & $80\pm11$    & 19   & S~II   \\
$1253.807\pm0.011$ &  Galactic       & $127\pm12$   & 28   & S~II   \\
                   &                 &              &      &        \\
$1259.519\pm0.007$ &  Galactic       &  $132\pm16$  & 15   & S~II        \\
$1259.869\pm0.019$ &  Galactic       &  $75\pm20$   & 8.8  & Si~II (HVC)$^d$ \\ 
$1260.398\pm0.055$ &  Galactic       &  $467\pm81$  & 53   & Si~II+Fe~II \\ 
$1270.802\pm0.014$ & $13,596\pm4$    &  $102\pm17$  & 12   & \Lya     \\ 
$1281.393\pm0.006$ & $16,208\pm2$    &  $345\pm22$  & 45   & \Lya     \\
$1281.867\pm0.063$ & $16,325\pm16$   &  $68\pm30$   & 9.2  & \Lya     \\
$1284.484\pm0.009$ & $16,970\pm2$    &  $448\pm22$  & 61   & \Lya     \\
$1285.097\pm0.013$ & $17,121\pm3$    &  $363\pm24$  & 48   & \Lya     \\
$1287.515\pm0.010$ & $17,717\pm3$    &  $141\pm16$  & 19   & \Lya     \\
$1288.976\pm0.020$ & $18,078\pm5$    &  $100\pm19$  & 13   & \Lya     \\
\hline \\
\end{tabular}
\end{center}

\noindent
$^a$ Top group of lines measured from previously unpublished,
pre-COSTAR GHRS/G160M spectrum (Fig. 2).  Bottom group is from our
post-COSTAR, GHRS/G160M spectrum (Fig. 1).  We use global continuum fits
and quote rest-frame equivalent widths (EW). 

\noindent
$^b$ LSR velocity ($cz$), which is the same as heliocentric
velocity for this sightline to within $\pm1$ \kms.   
Wavelength scales have been aligned 
assuming that the Galactic S~II lines lie at $V_{\rm LSR}$. This
procedure gives offsets of +0.029~\AA\ (new data) and
$-0.001$~\AA\ (pre-COSTAR data).  

\noindent
$^c$ Significance of the line (in $\sigma$) is defined as the integrated 
S/N per resolution element of the fitted absorption feature.
 
\noindent
$^d$ High velocity cloud was detected in C~IV at
$V_{\rm LSR} = -141 \pm 9$ \kms\ (Sembach \etal 1998).

\newpage
 
\begin{center}
{\bf Table 2} \\
{\bf VLA Instrumental Parameters } \\
\   \\
\begin{tabular}{lc}
\hline\hline
      &            \\
\hline
Date  & 1996 May \\
Configuration & DnC \\
Integration time (hrs) & 40 \\
Central Velocity (\kms) & 16,880 \\
Channel Width (kHz)& 195 \\
Channels & 31\\
Velocity Resolution (\kms) & 46 \\
Usable Velocity Range (\kms) & 16,283 - 17,571 \\
Synthesized beam (arcsec) & $54.4 \times 36.6$ \\
rms noise (mJy/beam) & 0.15 \\
rms column density sensitivity (cm$^{-2}$) & $4 \times 10^{18}$ \\
\hline \\
\end{tabular}
\end{center}

\begin{center}
{\bf Table 3} \\
{\bf H~I Properties of Detected Galaxies } \\
\   \\
\begin{tabular}{lcccccc}
\hline\hline
      &          &            &               &          & \\
Name & RA (1950) & Dec (1950) & $V_{\rm hel}$ & Velocity~Range & $M_{\rm HI}$ &
                              $L_B/L_*$ \\
     &           &            & (\kms)   & (\kms) & ($10^9~M_{\odot}$) 
                                   & \\ 
\hline
 
2155-3033   & 21 55 46.9  & -30 33 49.6 & 17,087 & 16,888 - 17,294 & 7.4
                                                                & 1.4  \\
F21569-3030 & 21 56 56.0  & -30 30 43.7 & 16,785 & 16,650 - 16,926 & 14 
                                                                & 0.9 \\
uncataloged & 21 56 03.6  & -30 17 08.7 & 17,179 & 17,064  -17,294 & 4.9
                                                                & 0.3  \\
APMBGC 466-053+024 & 21 55 29.9 & -30 33 52.6 &$\sim$ 16,200&$<$16,237
                               -16,375 & $>$ 3.6 & 3.0  \\
uncataloged & 21 56 38.5 & -30 30 57.4 & 17,156 & 17,156    & 0.8  & 0.014 \\
\hline \\
\end{tabular}
\end{center}

\newpage
 
\begin{center}
{\bf Table 4} \\
{\bf Predicted Column Densities$^a$ and Equivalent Widths} \\
\   \\
\begin{tabular}{ccccc}
\hline\hline
Ion   & Column Density       & $W_{\lambda}$ & $\lambda_0$ &  Observed
$\lambda_i$\\
      & (cm$^{-2}$)          & (m\AA)        &    (\AA)    &
[$\lambda_0(1+z_i)]$\\
\hline
      &                      &               &             &    \\
C~II  & $1.1 \times 10^{12}$ & 2.2           & 1334.532    &
1406.68,~1410.07,~1410.75 \\
C~III & $2.4 \times 10^{13}$ & 156           &  977.020    &
1029.84,~1032.32,~1032.82  \\
C~IV  & $1.8 \times 10^{13}$ &  73           & 1548.195    &
1631.90,~1635.83,~1636.61 \\
N~V   & $1.7 \times 10^{12}$ &  3.6          & 1238.821    &
1305.80,~1308.94,~1309.57 \\
O~VI  & $1.6 \times 10^{12}$ & 2.0           & 1031.926    &
1087.72, 1090.34,~1090.86  \\
Si~II & $6.5 \times 10^{10}$ & 1.0           & 1260.422    &
1328.56, 1331.77,~1332.40 \\
Si~III& $1.0 \times 10^{12}$ &  22           & 1206.500    &
1271.73,~1274.79,~1275.40  \\
Si~IV & $5.8 \times 10^{11}$ & 4.5           & 1393.755    &
1469.11,~1472.65,~1473.35 \\
\hline \\
\end{tabular}
\end{center}
 
\noindent
$^a$ Column densities are computed for cloud with N$_{\rm HI} =
2 \times 10^{16}$ cm$^{-2}$, $D = 400$ kpc, and 0.003 solar
metallicity, irradiated by QSO ionizing spectrum with $\alpha_b = 1.8$ 
and $I_0 = 10^{-23}$ ergs cm$^{-2}$ s$^{-1}$ Hz$^{-1}$ sr$^{-1}$. 
Observed wavelengths correspond to \Lya\ components at 1281.393, 1284.484,
and 1285.097 \AA, at redshifts $z_1 = 0.054063$, $z_2 = 0.056605$,
and $z_3 = 0.057110$.   Equivalent widths are quoted in the rest frame.  \\


\newpage

\begin{references}

 Allen, R. G., Smith, P. S., Angel, J. R. P., Miller, B. W.,
Anderson, S. F., \& Margon, B.  1993, ApJ, 403, 610 

 Appenzeller, I., Mandel, H., Krautter, J., Bowyer, S., Hurwitz, M.,
Grewing, M., Kramer, G., \& Kappelmann, N. 1995, ApJ, 439, L33
 
 Bahcall, J. N. et al. 1991, ApJ, 377, L5
 
 Bruhweiler, F. C., Boggess, A., Norman, D. J., Grady, C. A.,
Urry, C. M., \& Kondo, Y. 1993, ApJ, 409, 199

 Burles, S., \& Tytler, D.  1997, preprint (astro-ph/9712265) 

 Canizares, C. R., \& Kruper, J. 1984, ApJ, 278, L99

 Cen, R., Miralda-Escud\'e, J., Ostriker, J. P., \& Rauch, M. 1994,
ApJ, 437, L9
 
 Chen, H.-W., Lanzetta, K. M., Webb, J. K., \& Barcons, X. 1998,
    ApJ, 498, 77 

 Cowie, L. L., Songaila, A., Kim, T.-S., \& Hu, E. M.  1995, AJ, 109,
1522
 
 Donahue, M., \& Shull, J. M. 1991, ApJ, 383, 511

 Dove, J. B., \& Shull, J. M. 1994, ApJ, 423, 196

 Falomo, R., Pesce, J. E., \& Treves, A.  1993, ApJ, 411, L63 

 Ferland, G.  1996, Hazy, University of Kentucky Internal Report
  (Version 90.03)
 
 Giroux, M. L., \& Shull, J. M.  1997, AJ, 113, 1505

 Grevesse, N., \& Anders, E. 1989, in Cosmic Abundances of Matter,
AIP Conf. 183, ed. C. J. Waddington, 1. 

 Grogin, N. A., \& Geller, M. J. 1998, ApJ, in press (astro-ph/9804326) 
 
 Hernquist, L., Katz, N., Weinberg, D. H., \& Miralda-Escud\'e, J. 1996,
ApJ, 457, L51

 Hurwitz, M., et al. 1998, ApJ, 500, L61 

Lauberts, F., \& Valentijn, E. A. 1989, The Surface Photometry
Catalog of the ESO-Uppsala Galaxies, (Garching, ESO)
 
 Linsky, J. L., et al. 1993, ApJ, 402, 694 

 Linsky, J. L., et al. 1995, ApJ, 451, L335 

 Loveday, J. 1996, MNRAS, 278, 1025

 Madejski, G., et al. 1998, preprint, submitted to ApJ
 
 Maraschi, L., Blades, J. C., Calanchi, C., Tanzi, E. G., \& Treves, A.
1988, ApJ, 333, 660

 Marzke, R. O., Huchra, J. P., \& Geller, M. J. 1994, ApJ, 428, 43 
 
 Morris, S., Weymann, R. J., Savage, B., \& Gilliland, R. L. 1991,
ApJ, 377, L21

 Morris, S., Weymann, R. J., Dressler, A., McCarthy, P. J., Smith, B. A., 
Terrile, R. J., Giovanelli, R., \& Irwin, M. 1993, ApJ, 419, 524 

 Mulchaey, J. S., \& Zabludoff, A. I. 1998, ApJ, 496, 73 

 Mushotzky, R., \& Loewenstein, M. 1997, ApJ, 481, L63
 
 Penton, S., Shull, J. M., \& Edelson, R. 1998, Database of IUE-AGN 
Ultraviolet Spectra, available on Website 
(http://casa.colorado.edu/$\sim$spenton/IUEAGN/FUSE.html)
 
 Penton, S., Stocke, J. T., \& Shull, J. M. 1998, in preparation.

 Sandage, A. 1975, in Galaxies and the Universe, ed. by A. Sandage, M.
    Sandage, J. Kristian (University of Chicago Press, Chicago).
 
 Schramm, D. N., \& Turner, M.  1998, Rev. Mod. Phys., 70, 303

 Sembach, K. R., Savage, B. D., Lu, L., \& Murphy, E. M. 1998,
   ApJ, in press 

 Sherbert, L. E., \& Hulbert, S. J. 1997, GHRS Instrument Science Report 067,
Official Update (July 1997)

 Shull, J. M. 1997, in Structure and Evolution of the IGM
  from QSO Absorption Lines,  ed. P. Petitjean \& S. Charlot, 
(Paris, Editions Fronti\`eres),  101 

 Shull, J. M., Stocke, J. T., \& Penton, S.  1996, AJ, 111, 72

 Shull, J. M., Roberts, D., Giroux, M., \& Penton, S. 1998, in
   preparation. 
 
 Songaila, A. A. 1997, in Structure and Evolution of the IGM
  from QSO Absorption Lines, ed. P. Petitjean \& S. Charlot, 
 (Paris, Editions Fronti\`eres),  339 

 Songaila, A., \& Cowie, L. L. 1996, AJ, 112, 335
 
 Stark, A. A., Gammie, C. F., Wilson, R. W., Bally, J., Linke, R. A.,
   Heiles, C., \& Hurwitz, M.  1992, ApJS, 79, 77

 Steidel, C. C.,  1995, in QSO Absorption Lines, Proc. of ESO
Symposium, ed. G. Meylan (Heidelberg, Springer), 139

 Stocke, J., Shull, J. M., Penton, S., Donahue, M., \& Carilli, C.
1995, ApJ, 451, 24
 
 Tytler, D.  1995, in QSO Absorption Lines, Proc. of ESO
   Symposium, ed. G. Meylan (Heidelberg, Springer), 289

 van Gorkom, J. H. 1993, in The Environment and Evolution of Galaxies,
  ed. J. M. Shull \& H. A. Thronson, (Dordrecht, Kluwer), 345 

 van Gorkom, J. H., Carilli, C. L., Stocke, J. T., Perlman, E. S., \&
 Shull, J. M. 1996, AJ, 112, 1397

 Weymann, R. J., Rauch, M., Williams, R., Morris, S., \& Heap, S.
    1995, ApJ, 438, 650

 Wurtz, R., Stocke, J. T., Ellingson, E., \& Yee, H. K. C.  1997, ApJ, 
   480, 547

 Zabludoff, A. I., \& Mulchaey, J. S.,  1998, ApJ, 496, 39  

 Zheng, W., Kriss, G. A., Telfer, R. C., Grimes, J. P., \&
Davidsen, A. F. 1997, ApJ, 475, 469

\end{references}
\end{document}